\documentclass[manuscript, screen, nonacm]{acmart}   

\setcopyright{none}
\renewcommand\footnotetextcopyrightpermission[1]{}
\makeatletter
\gdef\@acmBooktitle{}%
\gdef\@acmConference{}%
\gdef\@acmYear{}%
\gdef\@acmPrice{}%
\gdef\@acmDOI{}%
\gdef\@acmISBN{}%
\makeatother

\usepackage{amsmath,amsfonts}
\usepackage{algorithmic}
\usepackage{float}
\usepackage{graphicx}
\usepackage{textcomp}
\usepackage{xcolor}
\usepackage{adjustbox}
\usepackage[font=small,labelfont=bf]{caption}
\usepackage[font=footnotesize,labelformat=simple]{subcaption}
\usepackage{subcaption}
\usepackage{bm}
\usepackage{comment}
\usepackage{subcaption}
\usepackage{epstopdf}
\usepackage{siunitx}
\usepackage{setspace}
\usepackage{array}
\usepackage{listings,multicol}
\usepackage{multirow}
\usepackage{lscape}
\usepackage{rotating}
\usepackage{xcolor, colortbl}
\usepackage{siunitx}
\usepackage{mathtools}
\usepackage{paralist}
\usepackage{todonotes}
\usepackage{wrapfig}
\usepackage{tabularx}
\usepackage[ruled,vlined,linesnumbered]{algorithm2e}

\usepackage{amsthm}
\theoremstyle{acmdefinition}

\usepackage[utf8]{inputenc}
\usepackage{tikz}
\usepackage[font={small}]{caption}
\usepackage[normalem]{ulem}

\newcolumntype{C}[1]{>{\centering}m{#1}}
\newcolumntype{L}[1]{>{\arraybackslash}m{#1}}
\newcolumntype{R}[1]{>{\RaggedLeft\arraybackslash}p{#1}}

\newcommand{\rowcol}{\rowcolor{black!5}}

\newif\ifTDraft
\TDraftfalse
\newcommand{\skipClean}[1]{}

\usepackage{caption}            
\renewcommand{\arraystretch}{1}

\AtBeginDocument{%
	}

\begin{document}

\title{\textit{FORSLICE}: An Automated Formal Framework for Efficient PRB-Allocation towards Slicing Multiple Network Services}

\author{Debarpita Banerjee}
\authornote{This is the corresponding author}
\email{debarpita2023_r@isical.ac.in}
\orcid{0009-0009-6193-4716}
\affiliation{
	\institution{Indian Statistical Institute Kolkata}
    \city{Kolkata}
	\country{India}
}

\author{Sumana Ghosh}
\email{sumana@isical.ac.in}
\orcid{0000-0002-5999-3313}
\affiliation{
	\institution{Indian Statistical Institute Kolkata}
    \city{Kolkata}
	\country{India}
}

\author{Snigdha Das}
\email{snigdha.das@ericsson.com}
\orcid{0000-0001-5801-7196}
\affiliation{
	\institution{Ericsson Research}
	\country{India}
}

\author{Shilpa Budhkar}
\email{shilpa.budhkar@ericsson.com}
\affiliation{
	\institution{Ericsson Research}
	\country{India}
}

\author{Rana Pratap Sircar}
\email{rana.pratap.sircar@ericsson.com}
\affiliation{
	\institution{Ericsson Research}
	\country{India}
}

\renewcommand{\shortauthors}{D. Banerjee et al.}

\begin{abstract}
\textit{Network slicing} is a modern 5G technology that provides efficient network experience for diverse use cases. It is a technique for partitioning a single physical network infrastructure into multiple virtual networks, called {\it slices}, each equipped for specific services and requirements. 
In this work, we particularly deal with {\it radio access network (RAN) slicing} and resource allocation to RAN slices. 
In 5G, {\it physical resource blocks (PRBs)} being the fundamental units of radio resources, our main focus is to allocate PRBs to the slices efficiently. 
While addressing a spectrum of needs for multiple services or the same services with multi-priorities, we need to ensure two vital system properties: i) {\it fairness} to every service type (i.e., providing the required resources and a desired range of throughput) even after prioritizing a particular service type, and ii) {\it PRB-optimality} or minimizing the unused PRBs in slices. These serve as the core performance evaluation metrics for PRB-allocation to RAN slices, in our work.

We adopt the \textit{3-layered hierarchical PRB-partitioning} technique for allocating PRBs to network slices.
The case-specific, AI-based solution of the state-of-the-art method lacks sufficient correctness to ensure consistent system performance.
To achieve guaranteed correctness and completeness, we leverage \textit{formal methods} and propose the first approach for a fair and optimal PRB distribution to RAN slices. 
We formally model the PRB-allocation problem as a {\it 3-layered framework}, {\it FORSLICE}, specifically by employing {\em satisfiability modulo theories}. Next, we apply formal verification to ensure that the desired system properties: fairness and PRB-optimality, are satisfied by the model. 
The proposed method offers an efficient, versatile and automated approach compatible with all \textit{3-layered hierarchical network structure} configurations, yielding significant system property improvements compared to the baseline.
\end{abstract}

\keywords{RAN Slicing, PRB-Allocation, Network Performance, Formal Verification, Satisfiability Modulo Theories}


\maketitle
\section{Introduction}
\label{sec:intro}
 
In telecommunication, {\it network slicing} is a technology that allows a shared physical network infrastructure to support multiple, diverse and service-specific virtual networks, called {\it slices}.
A network slice functions as an independent, secure, and fully operational network environment with its own {\it service level agreements (SLAs)}.
It can be customized to serve different essential system parameters like speed, latency, throughput, reliability, etc., depending on the needs of specific services or user groups. For instance, a network slice designed for IoT devices might prioritize high latency, while another slice for streaming video might focus on high bandwidth.

In this work, we consider {\it radio access network (RAN) slicing} and specifically focus on the aspect of radio-resource allocation to RAN slices. The fundamental unit of radio resources, available for data transmission, is a {\it physical resource block} or PRB. {\it PRB-allocation} is an elemental aspect of RAN resource scheduling, which distributes resources (divided into resource blocks) to user equipment (UEs). This work aims to obtain an efficient PRB-allocation strategy for a RAN slicing scenario.

In 5G networks and beyond, PRBs need to be distributed among multiple RAN slices corresponding to various services such as, {\it ultra-reliable low-latency communications (URLLC), enhanced mobile broadband (eMBB), massive machine-type communications (mMTC), fixed wireless access (FWA)}, etc., and also {\em multiple priorities} of a single service, e.g., {\em eMBB Premium} and {\em eMBB Normal} under the eMBB service. Hence, efficiently allocating PRBs is vital to meet {\it differentiated quality of service (QoS)} requirements (i.e., the ability to meet the service demands of different types of traffic).

A substantial body of existing literature addresses PRB-allocation and RAN slicing in several contexts, such as, media access control layer scheduling perspective~\cite{yin2020multiplexing,bakri2019dynamic}, resource allocation in remote radio heads~\cite{setayesh2020jointScheduling}, virtual network function activation~\cite{motalleb2022resource}, etc. However, our work addresses a  different context --- the \textit{PRB-partitioning} technique for allocating PRBs to RAN slices by considering a \textit{3-layered hierarchical network structure} of slices and partitions 
(RAN is divided into physical partitions, which are further sub-divided into logical units or slices).
This is \textit{Ericsson's} approach to RAN slicing~\cite{ericsson_RAN_slicing} which we follow in this work.
To our knowledge, PRB-allocation in such a context remains mostly unexplored in the literature, except the one studied in \cite{convergence}. We show that our proposed PRB-allocation approach is robust and more efficient as compared to the baseline~\cite{convergence} despite adopting the same PRB-partitioning technique~\cite{ericsson_RAN_slicing}.

\paragraph{\textbf{Motivation}}
In the context of PRB-partitioning, to cope with the diverse needs for {\em multi-services} or {\em the same service with multi-priorities}, there must be fair allocations of PRBs to different RAN slices. But at the same time, the optimal distribution of the PRBs among the multi-service-based slices is also a need, especially in the resource-constrained scenarios. A \textit{dependable} design strategy for PRB-allocation to RAN slices focuses on guaranteeing system properties, like ensuring fairness to each service type, optimal PRB distribution, meeting resource demands for higher priority service types, etc. These are essential factors to be addressed to uphold consistent network performance. The literature specifically lacks a dependable, generic PRB-allocation approach that takes as input any service-partition-slice configuration of the 3-layered hierarchical network structure, and efficiently allocates PRBs to partitions and slices while ensuring the system properties.

Although the method `\textit{Convergence}' proposed in \cite{convergence} explores PRB-allocation to RAN slices in the 3-layered hierarchical network structure, yet, their AI-planning based solution method is confined to a single case study, generating PRB-allocations only for a fixed configuration of service types, partitions, and slices. Moreover, the AI-based solution  of `\textit{Convergence}' fails to guarantee an acceptable degree of correctness, thus, not sufficient to ensure the continuity of the system performance in the deployment phase all the time.

In contrast, to address the issue of correctness, we leverage {\it formal methods} \cite{Formal_Clarke}, a well-proven mathematical technique for specifying and developing correct-by-construction designs of hardware and software systems. It not only verifies and ensures the correct functionalities of the underlying system but also ensures its completeness. 
 
Consequently, there is a recent trend of using formal methods in various industrial applications, such as, Internet of Things (IoT) \cite{Formal2023IoT}, cloud computing \cite{Formal2018cloud}, cyber-physical systems \cite{ghosh2020pattern,ghosh_19}, real-time scheduling~\cite{banerjee_FMSS}, design automation \cite{Dghosh2023BMC}, etc., over the last decade. Interestingly, in telecommunications, formal methods have been used in call admission control \cite{idi2022CAC}, 5G service orchestration \cite{UML_Uppaal}, secure transmission \cite{FV_EAP-AKA}, etc., but not in the domains of RAN slicing and PRB-allocation. 

Keeping the potential of correctness and completeness of any formal technique, we leverage {\it formal methods} to propose strategies for PRB-allocation to RAN slices in a 3-layered hierarchical network structure, which guarantees underlying system properties. The novelty of this work lies in being the first to apply formal methods to design an \textit{efficient, dependable PRB-allocation framework} that ensures system properties and provides \textit{differentiated QoS}.

\paragraph{\textbf{Overview of the Work}}
The two important {\em system properties} accounted for in our work are: {\em fairness} and {\em PRB-optimality}, functioning as the key network performance metrics in this work. Here, fairness particularly indicates the guarantee of providing the desired amount of PRBs and maintaining the throughput requirements for every service type, while PRB-optimality indicates minimized unused PRBs in slices. 
Furthermore, {\it best-effort services} (services having no specific QoS requirements like eMBB, FWA, etc.) are served with the unallocated PRBs (i.e., the unused spectrum or the PRBs not yet allocated to the slices).

Given a fixed total of available PRBs within a specific time interval, we incorporate user arrival data that follows the prescribed user distributions for different service types, as defined by 5G network protocols. The objective is to judiciously allocate PRBs to partitions and slices, thereby leaving sufficient unallocated resources to support best-effort services effectively.
However, fairness to all service types (like eMBB, FWA, URLLC, mMTC) must be guaranteed simultaneously, preserving the service type priorities. In this work, we prioritize the eMBB Premium service type above all others, guaranteeing its throughput requirements are met throughout runtime. 

To this end, we develop a {\it hierarchical $3$-layered framework}, \textit{FORSLICE}, that formally models the $3$-layered hierarchical network structure, and our goal is to ensure that the system properties are satisfied throughout the runtime. This formal framework is fully-automated to handle the automatic generation of PRB-allocations for any $3$-layered hierarchical network structure as input.

In particular, we use {\it satisfiability modulo theories (SMT), a popular} constraint-solving technique as the underlying formal method. We formulate a set of constraints for each layer modeling its functionality and working semantics. We also formalize the system properties. All the constraints are then fed to the SMT-solver for verification. A {\it satisfiable} answer returned by the solver provides an optimal PRB-allocation guaranteeing the system properties.

\paragraph{\textbf{Primary Contributions}}
This work contributes to the literature as follows.
\begin{enumerate}
    \item To our knowledge, for the first time, we leverage formal methods to propose an efficient, dependable PRB-allocation approach in the context of RAN slicing, ensuring system properties (fairness and PRB-optimality) while prioritizing the eMBB Premium service type and catering to best-effort services. 
    To accomplish this, we formally model the PRB-allocation problem as a 3-layered framework, \textit{FORSLICE}, which guarantees that all system properties remain valid throughout runtime.
    \item The proposed framework ensures generality and automated execution; it takes as input any service-partition-slice configuration of a 3-layered hierarchical network structure and automatically generates the PRB-allocation of slices and partitions, that ensures the system properties and preserves the service priority. As a result, \textit{FORSLICE} exhibits greater robustness than the baseline method \textit{`Convergence'}~\cite{convergence}, which is limited to a particular case study.
    \item The SMT-based simulations show that \textit{FORSLICE} ensures all the system properties, for each network configuration considered as input. Additionally, we present a case study discussion on network performance evaluation, to demonstrate that the throughput offered by the PRB-allocation in \textit{FORSLICE} matches the throughput observed in actual network simulations, while maintaining the priority to the eMBB Premium service type.
    \item Although the baseline \textit{`Convergence'}~\cite{convergence} addresses fair PRB distribution, the experimental comparisons show that our proposed method reports a $44.45\%$ improvement in system properties as compared to ~\cite{convergence}.
\end{enumerate}

\paragraph{\textbf{Organization}}
The paper is organized as follows. Section \ref{sec:RAN model} discusses the basic background of RAN slicing and system properties. Section \ref{sec:example} presents a case study discussion to highlight the motivation behind the proposed formal modeling. Section \ref{proposed_method} first elaborates on the constraint formulation in \textit{FORSLICE} for the case study, and then generalizes it for any input configuration.  
Section \ref{sec:results} presents experimental observations that justify the efficiency of the proposed framework. 
Section \ref{sec:network_simulation} discusses network performance evaluation scenarios.
Section \ref{sec:rwork} lists some related work and their limitations. Finally, Section \ref{sec:conclusion} concludes the entire work and its future directions.

\section{RAN Slicing Background}
\label{sec:RAN model}

We here describe the basic background of RAN slicing, PRB-allocation and system properties.

In general, network slicing is a technology that carves up a shared physical network into multiple logical networks, i.e., slices. To achieve RAN slicing, a common technique involves distribution and allocation of radio resources among different slices based on user requirements. A \textit{physical resource block (PRB)} is the unit of radio resource denoting the specific allocated spectrum block (defined as 12 consecutive subcarriers in the frequency domain). 

\begin{wrapfigure}{r}{0.51\linewidth}
    \centering
    \includegraphics[scale=0.25]{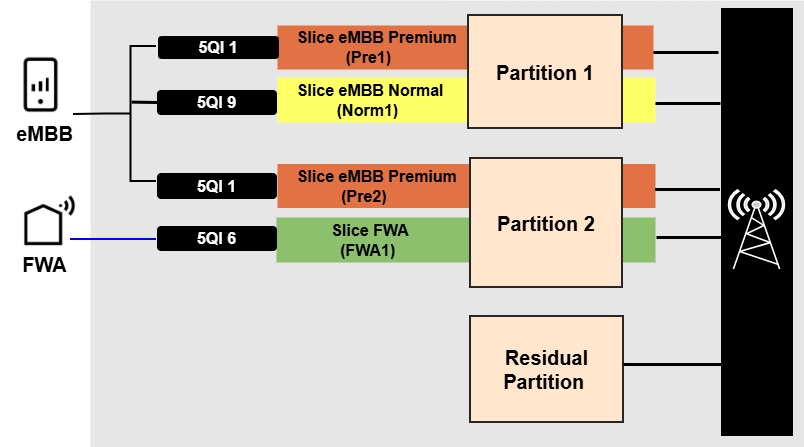}
    \caption{RAN slicing scenario [5QI: 5G quality identifier] }    
    \label{fig:RAN slicing}
\end{wrapfigure}

In this work, we follow \textit{Ericsson's} \textit{PRB-partitioning} technique~\cite{ericsson_RAN_slicing} for RAN slicing, that divides the RAN into physical \textit{partitions}, 
where a partition is a configured share of radio resources within a cell dedicated to specific user categories. 
Each partition is further comprised of logical or virtual segments called \textit{slices} based on the user category.
For example, an eMBB Premium slice is dedicated for 4K/8K video streaming applications.

The unallocated PRBs, i.e., the remaining PRBs, from the total available amount, which are not allocated to the partitions (i.e., basically its slices), are residual and form the \textit{residual partition}. This residual partition is mainly applicable for the {\it best-effort} services like background data transfers, email and messaging, etc., that have no specific QoS requirements like eMBB, FWA and others.

\begin{wrapfigure}{r}{0.52\linewidth}
    \centering
    \vspace{-2mm}
    \includegraphics[scale=0.18]{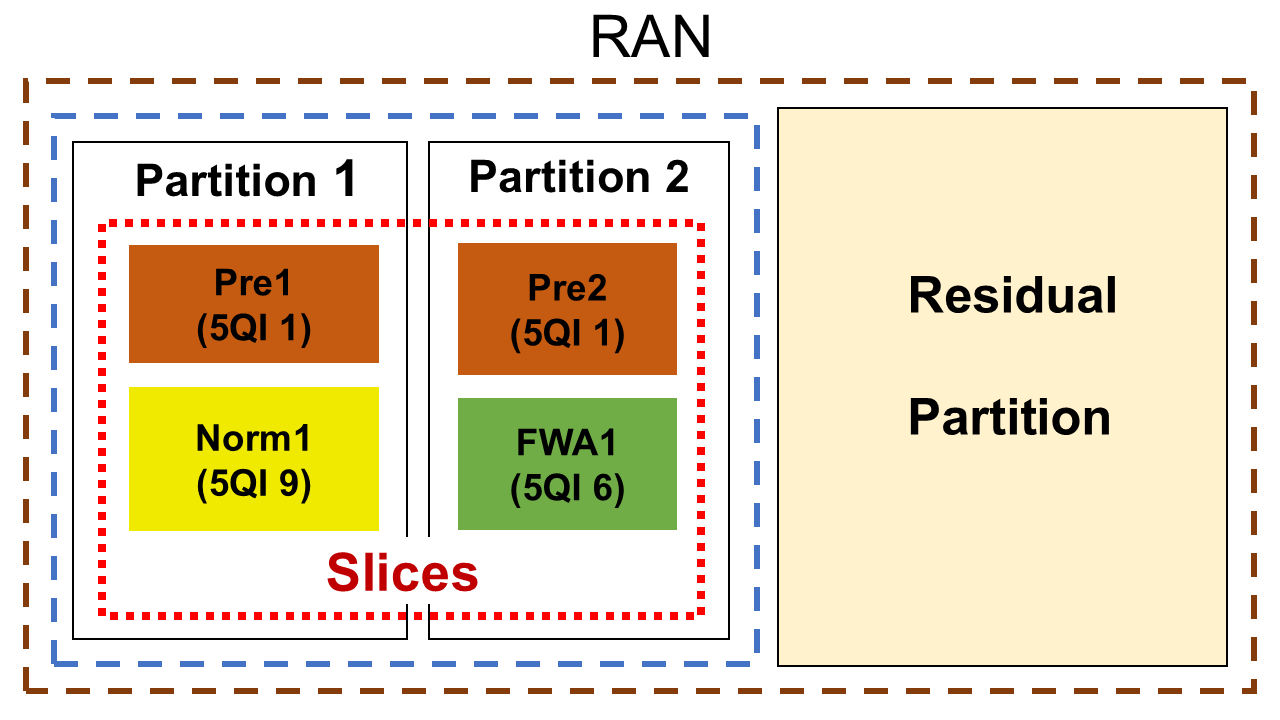}
    \vspace{-2mm}
    \caption{3-layered hierarchical network structure for the RAN-slicing scenario portrayed in Figure \ref{fig:RAN slicing}}   
    \label{fig:example_network_structure}
\end{wrapfigure}

Figure \ref{fig:RAN slicing} depicts an example scenario of PRB-partitioning and RAN slicing; where Partition 1 has $2$ slices: {\it eMBB Premium (Pre1)} and {\it eMBB Normal (Norm1)}, and Partition 2 has $2$ slices: {\it eMBB Premium (Pre2)} and {\it FWA (FWA1)}, and the rest is the residual partition.
On entering the network, a user is assigned to a slice based on its service type (e.g., eMBB Premium, eMBB Normal, etc.).

The hierarchical 3-layered network structure for this example is illustrated in Figure \ref{fig:example_network_structure}. The first layer is comprised of logical units or network slices; the upper layer includes the first layer along with two partitions; and the final or third layer supervises all the components of the RAN system: the first and second layers and the residual partition.

PRBs are dynamically allocated or de-allocated to or from slices to meet the resource demands of different service types.
For example, fair partitioning of PRBs involves allocating additional PRBs to an eMBB Normal slice when the number of users requesting that service is high in the day. Whereas, to optimize PRB usage, it necessitates to de-allocate some PRBs from a FWA slice on having a considerable low demand of FWA service during the late night hours.

Thus, in diverse service scenarios with varying QoS requirements, {\it fairness} and {\it PRB-optimality} are two vital {\em system properties} that need to be addressed while allocating PRBs to slices. Below, we define the notions of these properties that we consider in this work.
\begin{definition}[Fairness]
    \label{def:fairness}
    {\it Fairness is the guarantee of providing each service type with its desired amount of resources (i.e., PRBs) and throughput at all times, except when the residual partition is overused 
    (i.e., using up $x\%$ of total PRB-share)}. 
\end{definition}

\begin{definition}[PRB-optimality]
    \label{def:optimality}
    {\it PRB-optimality indicates minimizing unused PRBs  in the slices which are not used by the users assigned to the respective slices.}
\end{definition}

The two system properties described above exhibit opposing characteristics --- the former typically indicates towards allocating extra PRBs to slices to meet resource demands, whereas the latter suggests for de-allocation of PRBs for judicious usage of resources. The goal of our work is to ensure that the properties are simultaneously satisfied throughout runtime. `Minimizing unused PRBs' necessitates balancing optimality with fairness, indicating that a reduction or de-allocation of PRBs beyond the specified threshold, compromises the constraint of fairness.

For a given input configuration of partitions and slices, the proposed formal framework \textit{FORSLICE} determines a suitable PRB-allocation that preserves the system properties while prioritizing the eMBB Premium service type, thereby maintaining the desired network performance. Next, we consider an example scenario to present the motivation behind the proposed formal modeling.

\section{An Illustrative Example}
\label{sec:example}

Let us consider the example of the RAN-slicing and PRB-partitioning scenario shown in Figure \ref{fig:RAN slicing}. It deals with $3$ service types, viz., eMBB Premium, eMBB Normal and FWA; has $2$ partitions with two slices each: slices Pre1 and Norm1 in Partition $1$, and slices Pre2 and FWA1 in Partition $2$. 
The hierarchical 3-layered network structure for this example is shown in Figure \ref{fig:example_network_structure}. 

In this work, we present a 3-layered formal framework, \textit{FORSLICE}, which introduces various layer-specific techniques for efficient PRB-allocation in RAN-slicing scenarios. The rationale for this modeling approach is discussed here with reference to the above case study.

On entering the network, a user is assigned to a slice based on its service type. The primary processes executed in the slice layer (i.e., the first layer) after user assignment are as follows. There is an increase in the user count and PRB-usage with an increase in the number of users, and consequently, the remaining or residual amount of PRBs in the slice decreases. Users exiting the network trigger the reverse process. At any point of time, i) when there is an over-usage of PRBs and further requirement of extra resources, or ii) when an excess of PRBs remain residual, i.e., unutilized, for a long duration, some appropriate signals get generated in slices to indicate the needs for PRB-allocation and de-allocation, respectively.
We name the respective signals as \textit{top-up} and \textit{ramp-down} signals in our proposed method.

Any monitoring agent (in the upper or second layer) that controls and supervises over a partition and its slices, promptly takes the action for allocating or de-allocating PRBs to/from slices, based on the signals (top-up and ramp-down) generated. Since there are two slices (Pre1 and Norm1) in Partition 1, it may happen that at some point of time, both the slices require top-up (i.e., requirement of extra PRBs), or ramp-down (i.e., the need to de-allocate unused PRBs), or either of them requires a top-up and the other demands for a ramp-down, and so on.

The potential top-up and ramp-down scenarios for slices in Partition 1 are as follows: ($\phi_s$, $\phi_s$), ($\phi_s$, \texttt{TU}), (\texttt{TU}, $\phi_s$), ($\phi_s$, \texttt{RD}), (\texttt{RD}, $\phi_s$), (\texttt{TU}, \texttt{TU}), (\texttt{RD}, \texttt{RD}), (\texttt{TU}, \texttt{RD}) and (\texttt{RD}, \texttt{TU}). Here, the first and second elements of the tuple correspond to signal generation scenarios in slices Pre1 and Norm1 in Partition 1; and $\phi_s$, \texttt{TU} and \texttt{RD} refer to the scenarios when neither top-up nor ramp-down, only top-up, and only ramp-down signal is generated in a slice, respectively. Thus, each slice generates a signal in one of three ways at any given time: no activity (neither top-up nor ramp-down), only top-up, and only ramp-down. Simultaneous top-up and ramp-down in a single slice is not possible. For two slices in Partition 1, this results in a total of $3^2=9$ possible combinations of signal generations, to be handled by the monitoring agent of Partition 1. Similar arguments follow for Partition 2 as well. 

Based on such scenarios, the PRB-allocations or de-allocations have to be decided. For example, if both slices in Partition 1 require top-up, then PRBs need to be allocated to the slices, i.e., to Partition 1. However, in a scenario where Pre1 requests a top-up (say, of 10 PRBs) and Norm1 concurrently requests a ramp-down (to free 15 PRBs, say), an intra-partition adjustment can be made: allocate the 10 PRBs from the Norm1 slice to the Pre1 slice, and then de-allocate the remaining 5 PRBs from Partition 1. The PRB-allocations need to be adjusted accordingly for each of the nine distinct signal generation scenarios observed in Partition 1. Similar arguments also work for Partition 2.

On de-allocating PRBs from Partition 1 in the above example, an obvious intuition that occurs is that the de-allocated PRBs must be re-allocated to the residual partition. But the following scenario may also occur --- 5 PRBs are de-allocated from Partition 1 and simultaneously Partition 2 demands an allocation of extra 8 PRBs. 
Hence, a monitoring agent that supervises the entire system needs to decide on how to manage the residual partition's PRB-share. For example, it can design the following inter-partition adjustment: allocate 5 PRBs from Partition 1 to Partition 2; de-allocate 3 PRBs from the residual partition and allocate those to Partition 2. 

Since we have two partitions in this example, there are $3^2=9$ possible ways for PRB-adjustments: ($\phi_p$, $\phi_p$), ($\phi_p$, \texttt{A}), (\texttt{A}, $\phi_p$), ($\phi_p$, \texttt{D}), (\texttt{D}, $\phi_p$), (\texttt{A}, \texttt{A}), (\texttt{D}, \texttt{D}), (\texttt{A}, \texttt{D}) and (\texttt{D}, \texttt{A}). Here, the first and second elements of the tuple correspond to the allocation requirements in Partitions 1 and 2; and $\phi_p$, $\texttt{A}$ and $\texttt{D}$ refer to the scenarios when a partition requires neither allocation nor de-allocation, only allocation, and only de-allocation, respectively. 
Only one of these nine scenarios is possible at any given time; the necessary PRB-allocation and residual partition share adjustments must be implemented accordingly . 

Furthermore, the monitoring agent tasked with system supervision is also responsible for user assignment based on its service type, i.e., user-to-slice mapping. For example, each partition contains a Premium slice: Pre1 and Pre2 in Partitions 1 and 2 respectively. The monitoring agent hence needs to determine to which slice an eMBB Premium type user must be assigned after entering the network. If a random user assignment occurs, then that may lead to clustering of users and over usage of PRBs in one partition, thereby de-allocating excess PRBs from the residual partition. Hence, users must be proportionately assigned to slices --- assign a user to a slice within a partition such that it does not trigger excessive PRB-usage and a top-up scenario in the slice soon.

Our proposed strategies and user assignment policy address PRB-partitioning by ensuring fairness, PRB-optimality, and balanced use of the residual partition to benefit best-effort services while prioritizing the eMBB Premium service type over the others. The 3-layered formal framework \textit{FORSLICE} therefore, provides the formal modeling of the aforementioned strategies in the three layers that generates PRB-allocations while ensuring the system properties.
\section{The Design of FORSLICE}
\label{proposed_method} 
This work aims to develop a dependable RAN resource (i.e., PRB) allocation strategy that provides differentiated QoS guarantees. The design is engineered to achieve two main goals:
\begin{compactenum}[i)]
    \item Ensure fairness in PRB provisioning to all service types based on their specific demands.
    \item Minimize unused PRBs to serve the best-effort services accessing the residual-partition. Crucially, the eMBB Premium service type is designated the highest priority.
\end{compactenum}

\noindent
We leverage formal methods in our work to design such a dependable framework for a RAN slicing scenario, that efficiently meets the objectives.
As mentioned earlier, {\it FORSLICE} is 3-layered hierarchical framework where layers are modeled as follows.

\begin{enumerate}
    \item[\textbf{Layer 1:}] The lowest layer that models the {\it slices}.
    \item[\textbf{Layer 2:}] The  middle layer that models the {\it partitions} and their {\it monitoring agents}. These monitoring agents administer the partitions and their slices. 
    \item[\textbf{Layer 3:}] The topmost layer that models the {\it central monitoring agent}, monitoring over the entire system: slices, partitions and monitoring agents, as well as the residual partition.
\end{enumerate}

\begin{wrapfigure}{r}{0.5\linewidth}
    \centering
    \includegraphics[scale=0.2]{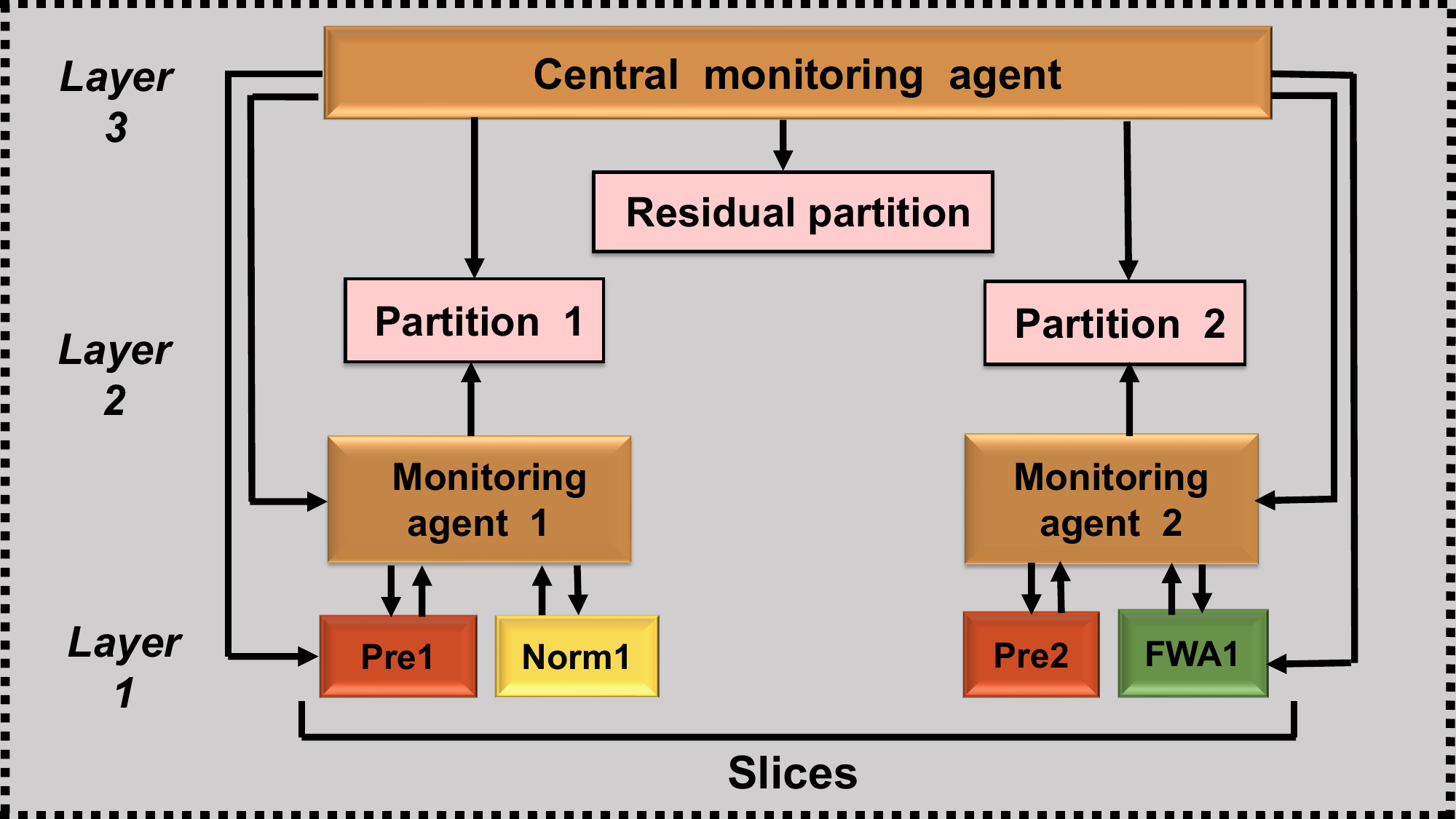}
    \caption{The proposed 3-layered framework, {\it \textbf{FORSLICE}}, for the sample example of \boldmath$3$ service types,  \boldmath$2$ partitions and \boldmath$2$ slices per partition}
    \label{fig:3-layered model}
\end{wrapfigure}

Figure \ref{fig:3-layered model} presents the 3-layered framework \textit{FORSLICE} for the example of $3$ service types, $2$ partitions and $2$ slices per partition, discussed in the previous section. Here, Layer $1$ consists of the four slices --- Pre1, Norm1, Pre2  and FWA1. 
Layer $2$ comprises two monitoring agents that monitor over Partitions $1$ and $2$ respectively, and also administer the slices in the respective partitions. Finally, the central monitoring agent in Layer $3$ manages the three layers and all the components, namely the slices, partitions, the monitoring agents and the residual partition. This indicates that the layer above governs the functions of all underlying layers.

\begin{wrapfigure}{r}{0.45\linewidth}
    \centering
    \includegraphics[scale=0.34]{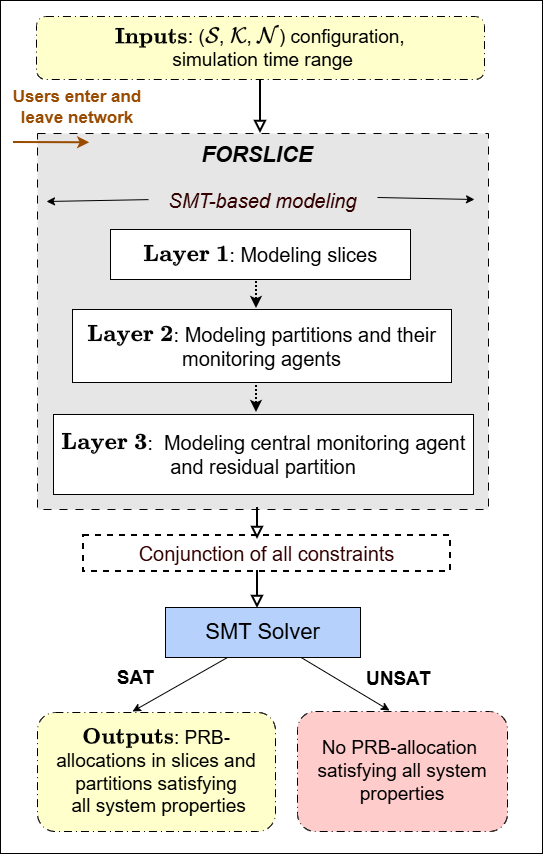}
    \caption{\textit{FORSLICE}: Automated design workflow}
    \vspace{-5mm}
    \label{fig:automated_workflow}
\end{wrapfigure}

Given any configuration of services ($\mathcal{S}$), partitions ($\mathcal{K}$) and slices ($\mathcal{N}$), \textit{FORSLICE} formally models these three layers and their properties one after another, starting from the lowest layer (i.e., Layer 1). Figure \ref{fig:automated_workflow} provides a pictorial representation of the workflow, which we have entirely automated. 

The ($\mathcal{S}$, $\mathcal{K}$, $\mathcal{N}$) configuration of service-partition-slice and the model simulation time range, are the inputs to our model. For example, let us suppose it requires to generate a fair and optimal PRB distribution during the busy hour, 9:00 AM-10:00 AM. Hence, we consider a fixed user distribution of services for the given time range based on the 5G telecommunication network protocols, like a heavy-tailed distribution for the eMBB service. The simulation time range being an input, the simulation begins with an initial state of zero active users.

In this work, we leverage SMT as the underlying formal technique for modeling and verification. A set of constraints is formulated during the sequential modeling of Layers 1, 2, and 3, using the provided inputs. The cumulative constraint --- the logical conjunction of all the individual constraints obtained in the three layers --- is supplied to the SMT solver for verification.

A \textit{satisfiable} or \texttt{SAT} answer from the solver indicates that \textit{FORSLICE} is able to generate PRB-allocations for the partitions and slices satisfying all the design objectives, i.e., ensuring the system properties and maintaining the priority to the eMBB Premium service type.

For the ease of presentation and feasibility of the SMT-based modeling, we develop a discretized version of the formal model. Therefore, the entire simulation on a time horizon of length $\hat{T}$ is discretized into multiple steps of length $h$ (say in the range of $\SI{1}{\minute}$), which means any discrete timestep $t$ signifies the actual time of $ht$ units. Thus, we simulate the model for $T=\lceil \frac{\hat{T}}{h} \rceil$ timesteps. 

Moreover, for modeling purposes, we make a key assumption: we scale the user count to a unit range in our model assuming transition of at most one user (i.e., one UE can enter or leave a slice) per timestep $t$. This simplification is essential for model abstraction, functioning similarly to how we might scale large data inputs (e.g., aggregating 1000 users into a single data point) to manage complexity.

For the sake of clarity and readability, we first delineate the entire formal modeling for the running example of $2$ partitions and $2$ slices per partition, discussed in Section \ref{sec:example}. Later, we generalize the  proposed modeling approach, described in Figure \ref{fig:automated_workflow}, for any given ($\mathcal{S}$, $\mathcal{K}$, $\mathcal{N}$) configuration of service-partition-slice as input.
For our running example, we first elaborate on all the variables and then the linear real arithmetic (LRA) constraints needed for the SMT-based formal modeling of each layer.

\subsection{Layer 1: Modeling the Slices}
\label{subsec:modeling slice}
We first present the structure of slices in Layer $1$. The variables used for formally modeling any slice {\it sl} are listed in Table \ref{tab:symbols_layer1}.
For simplicity and readability, we mostly refer to any slice as {\it sl} in the text and drop the subscript $i$ from $sl_{i,j}$. 

Initially, when the simulation begins (at timestep $j = 0$), there is no user entitled to the slice {\it sl}, leading to, {\it sl-usr$_{0}$ = 0, sl-usg$_{0}$ = 0} and {\it sl-resi$_{0}$= sl-shr$_{0}$}. \\

\textbf{Maximum possible usage of PRB over a time interval:}  
Given the time-window of slice {\it sl} as {\it sl-t-win}, different time intervals respecting {\it sl-t-win}, that we consider in our modeling, are {\it [((n-1) $\times$ sl-t-win) + 1, n $\times$ sl-t-win]}, for $n \in \mathbb{N}$. 
Without loss of generality, we assume that only one user can enter the slice at one timestep. Hence, at maximum {\it sl-t-win} many users can enter the slice during any such interval. 
Table \ref{tab:symbols_layer1} defines {\it sl-m} as the number of users who use up one PRB in slice $sl$.
This justifies the following relation,
\begin{equation}
\label{eq:max usage}
    maximum\ possible\ PRB{\text -}usage\ in\ an\ interval  =  \lceil \frac{sl{\text -}t{\text -}win}{sl{\text -}m} \rceil \ .
\end{equation}

The term `maximum' in the above equation reflects the fact that, this usage corresponds to the maximum number of possible users, {\it sl-t-win}, in any interval {\it [((n-1) $\times$ sl-t-win) + 1, n $\times$ sl-t-win]}.
Next, we define the following terminologies that help in managing the PRB-allocation and de-allocation in slices.

\begin{table}[b]
    \small \sf
    \renewcommand{\arraystretch}{1.2}
    \centering
    \begin{tabular}{|l|l|}
        \hline
        \textbf{Symbols} & \textbf{Meanings} \\ \hline
         $T$ & total timesteps derived from the simulation time horizon \\ \hline
         $sl{\text -}en_{i,j}$  & user entry flag (Boolean variable) for $i$-th slice at $j$-th timestep \\ \hline
         $sl{\text -}lv_{i,j}$  & user exit flag (Boolean variable) for $i$-th slice at $j$-th timestep \\ \hline
         $sl{\text -}usr_{i,j}$ & number of users in $i$-th slice at $j$-th timestep \\ \hline
         $sl{\text -}shr_{i,j}$ & number of PRBs allocated to $i$-th slice at $j$-th timestep \\ \hline
         $sl{\text -}usg_{i,j}$ & number of PRBs utilized by $i$-th slice at $j$-th timestep \\ \hline
         $sl{\text -}resi_{i,j}$ & number of residual PRBs (allocated but unused) in $i$-th slice at $j$-th timestep \\ \hline
         $sl{\text -}t{\text -}win_{i}$ &  {\begin{tabular}[c]{@{}c@{}}{time window of $i$-th slice i.e., no PRBs allocated or de-allocated to/from $i$-th} \\ {slice during interval $[((n-1) \times$ $sl{\text -}t{\text -}win_{i}) +1$, $n \times$ $sl{\text -}t{\text -}win_{i}$), $\forall n \in \mathbb{N}$} \end{tabular}} \\ \hline
         $sl{\text -}m_{i}$ & PRB-consumption of $i$-th slice: $sl{\text -}m_{i}$ users use up one PRB \\ \hline
         $sl{\text -}E_{i,j}$  &   {\begin{tabular}[c]{@{}c@{}}{number of users entered since the last time-window, in $i$-th} {slice at $j$-th timestep} \end{tabular}} \\ \hline
         $sl{\text -}top_{i,j}$ & top-up-signal flag (Boolean variable) in $i$-th slice at $j$-th timestep \\ \hline
         $sl{\text -}ramp_{i,j}$ & ramp-down-signal flag (Boolean variable) in $i$-th slice at $j$-th timestep \\ \hline
         $rp{\text -}ovr_{j}$ & residual partition-overuse flag (Boolean variable) at $j$-th timestep \\ \hline
    \end{tabular}
    \caption{Variables associated with Layer $1$}
    \label{tab:symbols_layer1}
\end{table}

\begin{definition} [Top-Up Signal] 
    \label{def:top up signal}
    {\it A top-up signal is generated in a slice if it requires additional PRBs.}
\end{definition}

\begin{definition} [Ramp-Down Signal]  
    \label{def:ramp down signal}
    {\it A ramp-down signal is generated in a slice to decrease the excess unused PRBs.}
\end{definition}

Since no PRB-allocation/de-allocation is allowed during intervals $[((n-1) \times sl{\text -}t{\text -}win) +1$, $n \times$ $sl{\text -}t{\text -}win)$, for all $n \in \mathbb{N}$ (see definition of {\it sl-t-win}$_{i}$ in Table \ref{tab:symbols_layer1}), thus, after every {\it sl-t-win} units, i.e., at $j=n \times sl{\text -}t{\text -}win$, $ n \in \mathbb{N}$,
it is checked whether slice {\it sl} requires a top-up or a ramp-down. Initially (at $j=0$), we set $sl{\text -}shr_{0} = \lceil \frac{sl{\text -}t{\text -}win}{sl{\text -}m} \rceil$, so that there are sufficiently many PRBs till the next window, following Eq. \eqref{eq:max usage}. 
During the formal modeling, we need to design the constraints for top-up and ramp-down signals in such a way, that a slice {\it sl} does not require any PRBs during the interval $[((n-1) \times sl{\text -}t{\text -}win) +1$, $n \times sl{\text -}t{\text -}win)$, for $n \in \mathbb{N}$.

A larger time interval, $[((n-1) \times sl{\text -}t{\text -}win) +1$, $n \times sl{\text -}t{\text -}win]$ of length $sl{\text -}t{\text -}win$ indicates a higher value of the maximum possible PRB-usage in that interval, which is $\lceil \frac{sl{\text -}t{\text -}win}{sl{\text -}m} \rceil$. 

{\it\textbf{ Inference 1:}} A larger time-window is desirable for the slices corresponding to service types with a  higher user-count and PRB-usage.

{\it \textbf{Inference 2:}} Conversely, a smaller time-window is preferable for the slices corresponding to the services with considerably fewer users. This leaves an option to frequently check whether the PRBs in the respective slices remain unused and if so, a ramp-down signal is generated soon to ensure PRB-optimality.

The PRB-consumption, {\it sl-m}, is chosen based on the resource demands of the service type of the slice {\it sl}. The eMBB Premium users have the highest priority, followed by eMBB Normal users, and then FWA users, in our running example. This prioritization ensures that eMBB Premium users consistently receive the required quality of service.

{\it \textbf{Inference 3:}} We consider a smaller PRB-consumption value for any slice corresponding to a higher priority service type, e,g., smaller PRB-consumption for eMBB Premium type indicates that the PRB-consumption by a single eMBB Premium user is higher as compared to other services. 

Let us consider an example. Usually there are more eMBB Normal type customers in comparison to the Premium type customers in the network, hence, we select the time-windows of the slices as, $sl{\text -}t{\text -}win> sl'{\text -}t{\text -}win$, where $sl$ and $sl'$ are any slices corresponding to the eMBB Premium and eMBB Normal service types respectively. 
Let $sl{\text -}m=2, sl{\text -}t{\text -}win=28$ and $sl'{\text -}m=3, sl'{\text -}t{\text -}win=40$. Therefore, we obtain, $\lceil \frac{sl{\text -}t{\text -}win}{sl{\text -}m} \rceil=14$ and $\lceil \frac{sl'{\text -}t{\text -}win}{sl'{\text -}m} \rceil=14$. Thus, the same number of PRBs, i.e., $14$ PRBs, are allocated to a Premium and to a Normal slice, but for intervals of lengths $28$ units and $40$ units respectively. This clearly infers that a larger number of PRBs are allocated to the Premium type customers over a smaller time interval, thereby, prioritizing them over other service types.

\textbf{Relationship between the throughput and PRB-usage:} The {\it key performance indicator (KPI)} that we consider in this work is \textit{throughput}. We desire to maintain a particular range of throughput for a particular service type (part of fairness, ref. Definition \ref{def:fairness}). 
We follow the equation given in \cite{3GPP} to calculate the maximum throughput, which is,

\vspace{-2.5mm}
\begin{equation*}
    T_h = 10^{-6} \ \sum_{j = 1}^J \ {\Large (} v^{(j)} \  Q_{m}^{(j)} \ f^{(j)} \ R_{max} \ \frac{N_{PRB}^{BW (j), \mu } \times 12}{T_{s}^{\mu}} \ ( 1 \ - \ OH^{(j)} ) \ {\Large )} \ .
\end{equation*}

Here, $J$ is the PRB-usage, $v^{(j)}$ is the maximum number of supported MIMO layers (separate data stream enabled by using multiple antennas at the base station), $Q_{m}^{(j)}$ is the maximum supported modulation order (the number of distinct symbols a digital communication system uses to encode data), $f^{(j)}$ is the scaling factor, $\mu$ is numerology (the flexible scaling of orthogonal frequency-division multiplexing (OFDM) waveform parameters), $T_{s}^{\mu}$ is the average OFDM symbol duration in a subframe for numerology $\mu$, $N_{PRB}^{BW (j), \mu}$ is the maximum resource block allocation in bandwidth $BW$ with numerology $\mu$, $OH^j$ is the overhead and $R_{max}$ is a scalar.

The Ericsson-specific choice for the values of these parameters are as follows:  $v^{(j)}=8$, $Q_{m}^{(j)}=64$, $f^{(j)}=1$, $R_{max}= \frac{948}{1024}$, $\mu=1$, $N_{PRB}^{BW (j), \mu}=38$, $T_{s}^{\mu}= \frac{10^{-3}}{14 \times 2^{\mu}}$ and $OH^{(j)}=0.14$. We set the value of throughput as $\hat{T}_{h}=\frac{80 \times T_{h}}{100}$ (since $T_{h}$ is the maximum), hence, we obtain the following equation,

\vspace{-3.5mm}
\begin{equation}
    \label{eq: throughput}
    \hat{T}_{h} = 4163.798 \times J \ .
\end{equation}

This proves that the throughput and PRB-usage are constant multiples of each other. Definitions \ref{def:top up signal} and \ref{def:ramp down signal} indicate that the top-up or ramp-down signals occur based on the PRB-usage in a slice at that timestep. Using Eq. \eqref{eq: throughput}, we can also interpret this as following --- upon the arrival of new users within a slice, additional PRB provisioning is necessary to achieve target throughput levels.

{\it \textbf{Inference 4:}} It is equivalent to consider any one of the two parameters, PRB-usage or throughput, to formalize the conditions for top-up/ramp-down signals. Therefore, if the PRB-allocation in \textit{FORSLICE} ensures the underlying system properties, it analogously validates the network performance efficacy by maintaining the intended throughput range. 
 
Next, we present the constraint formulation of Layer 1 for our running example.  
In this work, we assume that the residual partition should have at least $50\%$ of the total PRBs; if that is violated, then it is a {\it residual partition-overuse} scenario. The $50\%$ limit particularly guarantees adequate resource availability to the best-effort services.

\subsection{Constraints For Layer 1}
\label{subsec:constraints layer 1}
We now formally write down all the constraints pertaining to Layer $1$. A slice and a timestep are indexed with $i$ and $j$ respectively, in all the constraints written below. \\

\noindent
(1) {\bf Total user-count update for a particular timestep:} 
We use two Boolean variables, $sl{\text -}en_{i,j}$ and $sl{\text -}lv_{i,j}$, 
to design this constraint. The variables are defined as follows.\\ 
i) $sl{\text -}en_{i,j}$ is set to {\it True} ($\mathbb{T}$) if a user enters $i$-th slice at $j$-th timestep, and is {\it False} ($\mathbb{F}$) otherwise. \\
ii) $sl{\text -}lv_{i,j}$ is set to {\it True} ($\mathbb{T}$) if a user leaves $i$-th slice at $j$-th timestep, and is {\it False} ($\mathbb{F}$) otherwise. 

The user leaves a slice under two scenarios: 
a) leaving the network, and b) switching the service types (e.g., a Premium user shifts to the Normal type when its dedicated plan is exhausted). The user-count update constraint for our running example of four slices is as follows.  

    {\small
    {\small \boldmath$L_{1,1}$:} {\it {\Large $\land_{i = 1}^{4}$  $ \land_{j = 1}^{T}$} }
 
     {\it {\large[} ( sl-en$_{i,j}=\mathbb{T}$  $\land$  sl-lv$_{i,j}=\mathbb{F}$    
   \ $\Rightarrow$ \ sl-usr$_{i,j}=$ sl-usr$_{i,j{\text -}1}+1$ )}  
    {\Large $\land$} {\it ( sl-en$_{i,j}=\mathbb{F}$ \ $\land$ \  sl-lv$_{i,j}=\mathbb{T}$    
   \ $\Rightarrow$ \ sl-usr$_{i,j}=$ sl-usr$_{i,j{\text -}1}-1$ )} 

    {\Large $\land$}  {\it ( sl-en$_{i,j}=\mathbb{T}$  $\land$  sl-lv$_{i,j}=\mathbb{T}$ 
    \ $\Rightarrow$ \ sl-usr$_{i,j}=$ sl-usr$_{i,j{\text -}1}$ ) }    
    {\Large $\land$} {\it ( sl-en$_{i,j}=\mathbb{F}$  $\land$  sl-lv$_{i,j}=\mathbb{F}$      
     \ $\Rightarrow$ \ sl-usr$_{i,j}=$ sl-usr$_{i,j{\text -}1}$ )  {\large]} } }
     \\

At any $j$-th timestep, at most one user can enter into and can leave from a slice. Hence, the
user-count is updated at the $j$-th timestep in the following two cases: 

I. it is incremented when a user enters but no user leaves ($sl{\text -}en_{i,j}$ = $\mathbb{T}$  $\land$  $sl{\text -}lv_{i,j}$ = $\mathbb{F}$ ) \ and 

II. it is decremented when a user leaves and no user enters ($sl{\text -}en_{i,j}$ = $\mathbb{F}$  $\land$  $sl{\text -}lv_{i,j}$ = $\mathbb{T}$ ). \\
The user-count remains the same as that in the previous timestep for the other two cases: 

III. when one user enters and another leaves ($sl{\text -}en_{i,j}$ = $\mathbb{T}$  $\land$  $sl{\text -}lv_{i,j}$ = $\mathbb{T}$ ) \ and 

IV. when users neither enter nor leave ($sl{\text -}en_{i,j}$ = $\mathbb{F}$  $\land$  $sl{\text -}lv_{i,j}$ = $\mathbb{F}$ ). \\

\noindent
(2) {\bf User-count update for a particular time interval:} The number of users who have entered the $i$-th  slice till timestep $j$, starting from timestep, $n \times sl{\text -}t{\text -}win_i$ ($n \in \mathbb{N}$), is denoted by $sl{\text -}E_{i,j}$. The integer variable {\it sl-E} counts the number of users entering the slice $sl$ during any time interval [(({\it n-1}) $\times \ sl{\text -}t{\text -}win_{i}$) + $1$, $n$ $\times \  sl{\text -}t{\text -}win_{i}$], for $n \in \mathbb{N}$. 

For each $n \in \mathbb{N}$, at timestep $j = (n \times sl{\text -}t{\text -}win_i) + 1$, i.e., when $j \equiv 1$ mod ($sl{\text -}t{\text -}win_i$), $sl{\text -}E_{i,j}$ is set to $1$ if any user enters the $i$-th slice, or else it is set to $0$. This is because, the variable $sl{\text -}E_{i,j}$ is only specific to the time interval, [(({\it n-1}) $\times \ sl{\text -}t{\text -}win_{i}$) + $1$, $n$ $\times \  sl{\text -}t{\text -}win_{i}$], for each $n \in \mathbb{N}$. At all other timesteps, $sl{\text -}E_{i,j}$ is incremented only if a user enters. This constraint is formulated as follows. 

{\small
    {\small \boldmath$L_{1,2}$:} {\it {\Large $\land_{i = 1}^{4}$  $ \land_{j = 1}^{T}$}   
    {\Large [} \ {\Large (} j $\equiv$ $1$ mod (sl-t-win$_{i}$) $\Rightarrow$  
    ( sl-en$_{i,j}=\mathbb{T}$ $\Rightarrow$  sl-E$_{i,j}=1$ ) $\land$  
    ( sl-en$_{i,j}=\mathbb{F}$ $\Rightarrow$  sl-E$_{i,j} = 0$ ) {\Large )} 

    \hspace{2mm} {\Large $\land$} {\Large (} j $\not\equiv$ $1$ mod (sl-t-win$_{i}$) $\Rightarrow$  
    ( sl-en$_{i,j}=\mathbb{T}$  $\Rightarrow$  sl-E$_{i,j}=$ sl-E$_{i,j{\text -}1}$ + 1 ) $\land$ 
    ( sl-en$_{i,j}=\mathbb{F}$  $\Rightarrow$  sl-E$_{i,j}=$ sl-E$_{i,j{\text -}1}$ ) {\Large )} {\Large ]}} } \\

\noindent
This variable helps to design the constraint for ramp-down signal generation ($L_{1,5}$ in Section \ref{subsec:constraints layer 1}). \\

\noindent
(3) {\bf PRB-usage and residual PRB-count update:} We consider that one PRB is used up by $sl{\text -}m_{i}$ users (ref. Table \ref{tab:symbols_layer1}).  
Precisely, at the $j$-th timestep, when a user enters (i.e., $sl{\text -}en_{i,j}=\mathbb{T}$) and no user leaves (i.e., $sl{\text -}lv_{i,j}=\mathbb{F}$) and $sl{\text -}usr_{i,j}$ $\equiv$ 1 mod ($sl{\text -}m_{i}$), then the PRB-usage ($sl{\text -}usg_{i,j}$) is incremented by $1$ and residual PRB count ($sl{\text -}resi_{i,j}$) is decremented by $1$.
For example, let $sl{\text -}m_{i}=3$. Thus, when the {\it 1-st, 4-th, 7-th} and so on, users enter the slice (i.e., $sl{\text -}usr_{i,j}$ $\equiv$ $1$ mod $3$), $sl{\text -}usg_{i,j}$ is incremented by $1$ and $sl{\text -}resi_{i,j}$ drops by $1$.

Similarly, if a user leaves (i.e., $sl{\text -}lv_{i,j}=\mathbb{T}$) and no user enters (i.e., $sl{\text -}en_{i,j}=\mathbb{F}$) and $sl{\text -}usr_{i,j}$ $\equiv$ 0 mod ($sl{\text -}m_{i}$), then $sl{\text -}usg_{i,j}$ is decremented by $1$ and $sl{\text -}resi_{i,j}$ is incremented by $1$.
For example, there are four users in the $i$-th slice at timestep $j=5$, and suppose at $j=6$, one user leaves. Thus, the second PRB allocated to the $4$-th user is made free and one PRB is sufficient for the first three users, as $sl{\text -}m_{i}=3$. Hence, when users leave and the user-count drops to $3$, $6$, $9$ and so on, one PRB is freed, hence, $sl{\text -}usg_{i,j}$ is decremented and $sl{\text -}resi_{i,j}$ is incremented by $1$.

The constraint written below describes these two cases. For the other cases, both the variables, $sl{\text -}usg_{i,j}$  and $sl{\text -}resi_{i,j}$ remain the same as that in the previous timestep.

{\small
{\small \boldmath$L_{1,3}$}: {\it {\Large $\land_{i = 1}^{4}$  $ \land_{j = 1}^{T}$}

{\Large[} {\large [} {\large (} sl-en$_{i,j}=\mathbb{T}$ \ $\land$ 
sl-lv$_{i,j}=\mathbb{F}$ \  $\land$ \  sl-usr$_{i,j}$ $\equiv$ 1 mod (sl-m$_{i}$) {\large )}
 $\Rightarrow$ {\large (} sl-usg$_{i,j}=$ sl-usg$_{i,j{\text -}1} + 1$  \ {$\land$} \
  sl-resi$_{i,j}=$ sl-resi$_{i,j{\text -}1}\ -$ 1 {\large )} {\large ]}

 {\Large $\land$} \ {\large [} {\large (} sl-en$_{i,j}=\mathbb{F}$ \ $\land$ 
 sl-lv$_{i,j}=\mathbb{T}$ \ $\land$ \ sl-usr$_{i,j}$ $\equiv$ 0 mod (sl-m$_{i}$) {\large )}
 $\Rightarrow$ \ {\large (} sl-usg$_{i,j}=$ sl-usg$_{i,j{\text -}1} \ -$ 1 \  $\land$ \
 sl-resi$_{i,j}=$ sl-resi$_{i,j{\text -}1} + 1$ {\large )} {\large ]} {\Large]} } }\\

 \noindent
(4) {\bf Top-up signal generation:} The requirement for top-up in the $i$-th slice is checked after every $sl{\text -}t{\text -}win_{i}$ time units, beginning at the timestep $j=sl{\text -}t{\text -}win_{i}$. If in any interval [(({\it n-1}) $\times$ $sl{\text -}t{\text -}win_{i}$) + $1$, $n$ $\times$ $sl{\text -}t{\text -}win_{i}$), $n \in \mathbb{N}$, there has been a considerable amount of PRB-usage and there is not enough PRBs till the next window, then a top-up signal is generated at $j = n \times sl{\text -}t{\text -}win_{i}$. 

This is checked by the \textit{condition}: whether the residual number of PRBs ($sl{\text -}resi_{i,j}$) in the slice is less than or equal to the maximum usage ($\lceil \frac{sl{\text -}t{\text -}win_{i}}{sl{\text -}m_{i}} \rceil$) in the next interval. If the condition is true, then a top-up signal is generated. 

Moreover, the residual partition should not be overused (ref. Section \ref{subsec:modeling slice}) before generating a top-up signal, otherwise allocation of additional PRBs is not possible. We use a Boolean variable $rp{\text -}ovr_{j}$ to indicate 
a residual partition-overuse scenario (using up $50\%$ of total PRBs).
At the $j$-th timestep, $rp{\text -}ovr_{j}$ is {\it True} if the residual partition has less than $50\%$ of the total PRBs, else it is {\it False}. 

The constraint below highlights the case when the top-up signal, $sl{\text -}top_{i,j}$, is {\it True}, and in all other cases, it is {\it False}.

{\small
{\small \boldmath$L_{1,4}$:} {\it {\large $\land_{i = 1}^{4}$  $ \land_{j = 1}^{T}$} 
{\Large[} {\large(} rp-ovr$_{j}=\mathbb{F}$ \ $\land$ \ j $\equiv$ 0 mod (sl-t-win$_{i}$) 
 $\land$ \ sl-resi$_{i,j}$ $\le$ $\lceil \frac{sl{\text -}t{\text -}win_{i}}{sl{\text -}m_{i}} \rceil$ {\large)} \ $\Rightarrow$ sl-top$_{i,j}=\mathbb{T}$ {\Large]} } }\\

\noindent
(5) {\bf Ramp-down signal generation:} Similar to top-up, the condition for ramp-down in the $i$-th slice is checked at every $sl{\text -}t{\text -}win_{i}$ time units. The \textit{condition} to be checked is --- if at some timestep, $j = n \times sl{\text -}t{\text -}win_{i}$, the number of residual PRBs is quite large in amount, even after considering that the maximum usage amount ($\lceil \frac{sl{\text -}t{\text -}win_{i}}{sl{\text -}m_{i}} \rceil$) is available till the next time-window, $j = (n+1) \times sl{\text -}t{\text -}win_{i}$,
then a ramp-down signal ($sl{\text -}ramp_{i,j}$) is generated at $j = n \times sl{\text -}t{\text -}win_{i}$. 

At the $j$-th timestep, this `quite large' is quantified by two conditions:

i) the residual number of PRBs ($sl{\text -}resi_{i,j}$) is larger than twice the maximum usage in a time-interval, i.e., there are more than $\lceil \frac{sl{\text -}t{\text -}win_i}{sl{\text -}m_i} \rceil$ amount of PRBs, even after subtracting $\lceil \frac{sl{\text -}t{\text -}win_i}{sl{\text -}m_i} \rceil$ from $sl{\text -}resi_{i,j}$ (i.e., the residual PRBs can be utilized for the next two time intervals and even more),

ii) no users have entered the $i$-th slice during the interval [(({\it n-1}) $\times$ $sl{\text -}t{\text -}win_{i}$) + $1$, $n$ $\times$ $sl{\text -}t{\text -}win_{i}$], $n \in \mathbb{N}$, checked by the constraint, $sl{\text -}E_{i,j}= 0$. \\
If these two conditions are true, this clearly indicates a requirement to reduce the unused PRBs through a ramp-down. The following constraint captures this idea.

{\small
{\small \boldmath$L_{1,5}$:} {\it {\large $\land_{i = 1}^{4}$  $ \land_{j = 1}^{T}$} 
{\Large[} {\large (} j $\equiv$ 0 mod (sl-t-win$_{i}$) \ $\land$ \
 sl-resi$_{i,j}$ $-\  \lceil \frac{sl{\text -}t{\text -}win_i}{sl{\text -}m_i} \rceil$  $\ge$ 
 $\lceil \frac{sl{\text -}t{\text -}win_i}{sl{\text -}m_i} \rceil$ 
$\land$ \ sl-E$_{i,j} = 0 $ {\large )} \ $\Rightarrow$ \ sl-ramp$_{i,j}=\mathbb{T}$ {\Large]} } }\\

(6) {\bf Conflict removal:} The constraint below describes that both top-up and ramp-down signals cannot be generated at the same timestep, in the $i$-th slice, $1 \le i \le 4$.

{\small
{\small \boldmath$L_{1,6}$:} {\it {\large $\land_{i = 1}^{4}$  $\land_{j = 1}^{T}$ } 
{\large[}sl-top$_{i,j}=\mathbb{T}$ $\Rightarrow$ sl-ramp$_{i,j}=\mathbb{F}$  {\small $\land$}  sl-ramp$_{i,j}=\mathbb{T}$ $\Rightarrow$ sl-top$_{i,j}$= $\mathbb{F}$ {\large]} } }
   
\subsection{Layer 2:  Modeling the Partitions}
\label{subsec:modeling monitoring agent}
Next, we proceed to formally model the partitions in Layer $2$. This layer consists of the monitoring agents. A monitoring agent monitors over a partition $P$ and all its slices (i.e., the set $S_{P}$), by updating the PRB-shares of both $P$ and $S_{P}$. The PRB-share of partition $P$ is the total number of PRBs allocated to the slices in $S_{P}$. A slice may generate a top-up or ramp-down signal at any of the timesteps, $n \times sl{\text -}t{\text -}win$, $n \in \mathbb{N}$. Since, in our running example, each of the two partitions has two slices, there may be a scenario that both the slices of the same partition generate top-up-signals together, or, one slice requires top-up and the other opts for ramp-down (see the discussion in Section \ref{sec:example}). A partition's monitoring agent adjusts the PRB-shares accordingly for all such cases. We next define some terminologies. 

\begin{definition} [Top-Up-Slice]
\label{def:slice top-up}
{\it A top-up-slice is an action that allocates extra PRBs to a slice and increases its PRB-share, when the slice has generated a top-up signal.}
\end{definition}

\begin{definition} [Ramp-Down-Slice]
\label{def:slice ramp-down}
{\it A ramp-down-slice is an action that de-allocates excess unused PRBs from a slice and decreases its PRB-share, when the slice has generated a ramp-down signal.}
\end{definition}

\begin{definition}[Top-Up-Partition] 
\label{def:Top up partition}
{\it A top-up-partition is an action which increases the PRB-share of a partition because a top-up-slice action is executed in at least one of its slices.}   
\end{definition}

\begin{definition}[Ramp-Down-Partition] 
\label{def:Ramp down partition}
{\it A ramp-down-partition is an action which decreases the PRB-share of a partition because a ramp-down-slice action is executed in at least one of its slices.}   
\end{definition}

\subsection{Constraints For Layer 2}
For the constraint formulation, we use indices $k = 1$ and $2$ to represent the two partitions in our running example. 
The two slices in Partition $1$ are, $sl_1$ (Pre1) and $sl_2$ (Norm1) and
the two slices in Partition $2$ are, $sl_3$ (Pre2) and $sl_4$ (FWA1). Most of the variables in Layer $1$ (ref. Table $1$) are used once again in Layer $2$ (since the monitoring agents in Layer $2$ also monitor over slices). The new variables introduced in Layer $2$ are listed in Table \ref{tab:symbols_layer2}.

\label{subsec:contraints layer 2}
\begin{table}[h]
    \small \sf
    \renewcommand{\arraystretch}{1.3}
    \centering
    \begin{tabular}{|l|l|}
        \hline
        \textbf{Symbols} & \textbf{Meaning} \\ \hline
       $pt{\text -}shr_{k,j}$  &  the PRB-share of  $k$-th partition at $j$-th timestep \\ \hline
        $\pi_k$ & {\begin{tabular}[c]{@{}c@{}} {slice indices in $k$-th partition, e.g., $\pi_1=\{1,2\},$ } {$\pi_2=\{3,4\}$, for our example} \end{tabular}} \\ \hline
        $\hat{W}_i$ & the maximum usage amount $\lceil \frac{sl{\text -}t{\text -}win_i}{sl{\text -}m_i} \rceil$ for $i$-th slice \\ \hline
    \end{tabular}
    \caption{New variables associated with Layer 2}
    \label{tab:symbols_layer2}
\end{table}

In case of top-up-slice or ramp-down-slice actions, the amount of PRBs added or deducted to/from a slice $sl$, equals the maximum usage amount, $\lceil \frac{sl{\text -}t{\text -}win}{sl{\text -}m} \rceil$, in any interval $[((n{\text -}1) \times sl{\text -}t{\text -}win)+1, n \times sl{\text -}t{\text -}win]$ (ref. Eq. \eqref{eq:max usage} in Section \ref{subsec:modeling slice}). After a top-up-slice action, this ensures that the PRB-share of slice $sl$ is sufficient till the next time-window, thereby, guaranteeing fairness. 
Likewise, for the ramp-down-slice action, de-allocating $\lceil \frac{sl{\text -}t{\text -}win}{sl{\text -}m} \rceil$ many PRBs ensures fairness, as well as PRB-optimality. 

The rationale behind this is explained as follows. The ramp-down signal generation constraint ($L_{1,5}$) in Layer $1$ mentions that a ramp-down signal is generated in $sl$ at timestep $j = n \times sl{\text -}t{\text -}win$, when no users have entered the slice $sl$ during the interval $[((n{\text -}1) \times sl{\text -}t{\text -}win)+1, n \times sl{\text -}t{\text -}win]$, and also the unused PRBs (\textit{sl-resi}) in $sl$ are in excess (equal or more than $2 \times  \lceil \frac{sl{\text -}t{\text -}win}{sl{\text -}m} \rceil$), i.e., there are sufficiently many PRBs to get through the next two time-windows. If \textit{sl-resi}$_{i,j}= 2 \times  \lceil \frac{sl{\text -}t{\text -}win_i}{sl{\text -}m_i} \rceil$ for the $i$-th slice at the $j$-th timestep, then de-allocating more than $\lceil \frac{sl{\text -}t{\text -}win_i}{sl{\text -}m_i} \rceil$ many PRBs from the slice after a ramp-down-slice action may jeopardize the constraint of fairness; since the remaining PRBs would not be sufficient to cope up with the user requirements till the next time window. Therefore, we de-allocate $\lceil \frac{sl{\text -}t{\text -}win_i}{sl{\text -}m_i} \rceil$ many PRBs after a ramp-down-slice action to ensure fairness and PRB-optimality simultaneously.

{\it \textbf{Inference 5:}} Allocating and de-allocating $\lceil \frac{sl{\text -}t{\text -}win}{sl{\text -}m} \rceil$ many PRBs after a top-up-slice  and ramp-down-slice action respectively, ensure that there are sufficiently many PRBs till the next window (fairness) and also minimizes the unused PRBs (PRB-optimality).

In all the constraints written below, we use $i,i' \in \pi_k$ (ref. Table \ref{tab:symbols_layer2}) to denote the two slices in the $k$-th partition and $j$ is used to denote the timestep. Also, the quantity $\lceil \frac{sl{\text -}t{\text -}win_i}{sl{\text -}m_i} \rceil$ (ref. Eq. \eqref{eq:max usage}) is referred to as $\hat{W}_i$ for the $i$-th slice in all the constraints, for simplicity. We now elaborate and formulate the constraint for all nine cases discussed in Section \ref{sec:example}. \\

\noindent
(1)  {\bf Neither top-up-slice nor ramp-down-slice action:}
None of the slices in each of the two partitions require top-up-slice or ramp-down-slice action, maintaining the same partition shares as in previous timesteps. The constraint for this (case $(\phi_s,\phi_s)$ in Section \ref{sec:example}) is as follows.

{\small
{\small \boldmath$L_{2,1}$:} {\it {\large $\land_{k = 1}^{2} \land_{j = 1}^{T}$}
{\Large[}{\large $\land_{i \in \pi_k }$} {\large (}sl-top$_{i,j}$ $=$ $\mathbb{F}$ $\land$ sl-ramp$_{i,j}$ $=$ $\mathbb{F}${\large )}
$\Rightarrow$ pt-shr$_{k,j}$ = pt-shr$_{k,j{\text -}1}${\Large]} } }\\

\noindent
(2) {\bf Top-up-slice action for any one slice:} 
Only one of the two slices requires top-up and the other opts for no changes (i.e., cases $(\phi_s,\texttt{TU})$ and $(\texttt{TU},\phi_s)$ in Section \ref{sec:example}). Hence, it gives rise to a top-up-partition action. For example, in Partition $1$, slice $sl_1$ generates a top-up signal but slice $sl_2$ doesn't generate any top-up or ramp-down signal at the $j$-th timestep, which is formally written as, {\small $sl{\text -}top_{1,j}=\mathbb{T} \ \land \ sl{\text -}top_{2,j}=\mathbb{F} \ \land \ sl{\text -}ramp_{2,j}=\mathbb{F}$}.

Hence, there will be a top-up-slice action that increases the PRB-share, $sl{\text -}shr_{1,j}$, and the residual PRB count, $sl{\text -}resi_{1,j}$, of $sl_1$ . As the total PRB-share of $sl_1$ is increased and currently the blocks are unused, hence the residual PRB count is also increased.
The top-up-slice action in $sl_1$ leads to a top-up-partition action in Partition $1$ (increasing the partition's PRB-share $pt{\text -}shr_{1,j}$). 
The PRB-share of Partition 2, i.e.,  $sl{\text -}shr_{2,j}$, and the residual PRB count, i.e., $sl{\text -}resi_{2,j}$, of slice $sl_2$, remain the same as that in the $(j{\text -}1)$-th timestep. 
The constraint is written as follows.

{\small
{\small \boldmath$L_{2,2}$:}  
{\it {\large $\land_{k = 1}^{2}$  $\land_{j = 1}^{T}$} {\large $\land_{\substack{i,i' \in \pi_k, \ i \neq i'}}$}

{\Large[} \ {\Large (}sl-top$_{i,j}=\mathbb{T}$ $\land$ sl-top$_{i',j}=\mathbb{F}$ $\land$ 
sl-ramp$_{i',j}=\mathbb{F}${\Large )}  
$\Rightarrow$  {\Large (} sl-shr$_{i,j}=$ sl-shr$_{i,j{\text -}1}$ $+$ $\hat{W}_i$ \ $\land$  
 \ sl-resi$_{i,j}=$ sl-resi$_{i,j{\text -}1}$ $+$ $\hat{W}_i$

\hspace{15mm}
$\land$ \ sl-shr$_{i',j}=$ sl-shr$_{i',j{\text -}1}$ \ $\land$ \ sl-resi$_{i',j}=$ sl-resi$_{i',j{\text -}1}$
\ $\land$ \ pt-shr$_{k,j}=$ pt-shr$_{k,j{\text -}1}$ $+$ $\hat{W}_i${\Large )} {\Large]} } } \\

\noindent
(3) {\bf Ramp-down-slice action for any one slice:} 
Only one of the two slices generates a ramp-down signal and the other opts for no changes (i.e., cases $(\phi_s,\texttt{RD})$ and $(\texttt{RD},\phi_s)$ in Section \ref{sec:example}), resulting in a ramp-down-partition action. The constraint formulation in this case is similar to that of the previous case. Here, PRB-shares of a partition and its corresponding slice are decreased.

{\small
{\small \boldmath$L_{2,3}$:}  {\it {\large $\land_{k = 1}^{2}$$\land_{j = 1}^{T}$$\land_{\substack{i,i' \in \pi_k, i \neq i'}}$} 

{\Large[} \ {\Large (}sl-ramp$_{i,j}$ $=$ $\mathbb{T}$ $\land$ sl-top$_{i',j}$ $=$ $\mathbb{F}$ $\land$ 
sl-ramp$_{i',j}$ $=$ $\mathbb{F}${\Large )}
$\Rightarrow$ {\Large (} sl-shr$_{i,j}$ $ =$ sl-shr$_{i,j{\text -}1}$ $-$ $\hat{W}_i$ \ $\land$  
\ sl-resi$_{i,j}$  $=$ sl-resi$_{i,j{\text -}1}$ $-$ $\hat{W}_i$

\hspace{15mm}
$\land$ \ sl-shr$_{i',j}$  $=$ sl-shr$_{i',j{\text -}1}$ \ $\land$ \ sl-resi$_{i',j}$  $=$ sl-resi$_{i',j{\text -}1}$ 
\ $\land$ \ pt-shr$_{k,j}$ $=$  pt-shr$_{k,j{\text -}1}$ $-$ $\hat{W}_i$ {\Large )} \ {\Large]} } }\\

\noindent
(4) {\bf Top-up-slice action for both the slices:} 
Both the slices of a partition generate top-up signals simultaneously (i.e., case $(\texttt{TU},\texttt{TU})$ in Section \ref{sec:example}), and thus initiate a top-up-partition action. For example, slices $sl_1$ and $sl_2$ of Partition $1$ (note that $\pi_1=\{1,2\}$) require top-up at the $j$-th timestep concurrently, which is denoted as, $sl{\text -}top_{1,j}=\mathbb{T} \land sl{\text -}top_{2,j}=\mathbb{T}$. The top-up slice action increases the PRB-share ($sl{\text -}shr_{i,j}$) as well as the residual PRB count ($sl{\text -}resi_{i,j}$) of the $i$-th slice, by an amount of $\hat{W}_i$ (for all $i \in \pi_k$). Moreover, the top-up-partition action increases the PRB-share of Partition $1$ ($pt{\text -}shr_{1,j}$) by the amount, $\lceil \frac{sl{\text -}t{\text -}win_{1}}{sl{\text -}m_{1}} \rceil + \lceil \frac{sl{\text -}t{\text -}win_{2}}{sl{\text -}m_{2}} \rceil = \hat{W}_1 + \hat{W}_2$. We write the constraint below.  

{\small
{\small \boldmath$L_{2,4}$:} {\it {\large $\land_{k = 1}^{2}$  $\land_{j = 1}^{T}$  } 
{\large $\land_{\substack{i,i' \in \pi_k, \ i < i'}}$} {\Large[} {\large (} sl-top$_{i,j}$ $=$ $\mathbb{T}$ \ $\land$ \ sl-top$_{i',j}$ $=$ $\mathbb{T}$ {\large )}   

$\Rightarrow$ {\Large (}
{\large $\land_{\substack{i \in \pi_k}}$} {\large (} sl-shr$_{i,j}$  $=$ sl-shr$_{i,j{\text -}1}$ $+$ $\hat{W}_i$ \ 
$\land$ \
sl-resi$_{i,j}$  $=$ sl-resi$_{i,j{\text -}1}$ $+$ $\hat{W}_i$ {\large )}
$\land$ {\large (} pt-shr$_{k,j}$ $=$ pt-shr$_{k,j{\text -}1}$ $+$ $\sum_{i \in \pi_k}$ $\hat{W}_i$
{\large )} {\Large )} {\Large]} } }\\

\noindent
(5) {\bf Ramp-down-slice action for both the slices:}
Both the slices generate ramp-down signals simultaneously (i.e., case $(\texttt{RD},\texttt{RD})$ in Section \ref{sec:example}) and thus, ramp-down-slice and ramp-down-partition actions occur.
The constraint formulation is similar to the previous case.

{\small
{\small \boldmath$L_{2,5}$:} {\it {\large $\land_{k = 1}^{2}$  $\land_{j = 1}^{T}$  } 
{\large $\land_{\substack{i,i' \in \pi_k, \ i < i'}}$} {\Large[} {\large (} sl-ramp$_{i,j}$ $=$ $\mathbb{T}$ \ $\land$ \ sl-ramp$_{i',j}$ $=$ $\mathbb{T}$ {\large )}   

$\Rightarrow$ {\Large (}
{\large $\land_{\substack{i \in \pi_k}}$} {\large (} sl-shr$_{i,j}$ $=$ sl-shr$_{i,j{\text -}1}$ $-$ $\hat{W}_i$  
 \ $\land$ 
\ sl-resi$_{i,j}$ $=$ sl-resi$_{i,j{\text -}1}$ $-$ $\hat{W}_i$ {\large )}
$\land$  {\large (} pt-shr$_{k,j}$ $=$ pt-shr$_{k,j{\text -}1}$ $-$ $\sum_{i \in \pi_k}$ $\hat{W}_i$
{\large )} {\Large )} {\Large]} } }\\

\noindent
(6)  {\bf Top-up-slice action in one slice and ramp-down-slice action in another:} Here, two slices of the same partition have conflicting requirements (i.e., cases $(\texttt{TU},\texttt{RD})$ and $(\texttt{RD},\texttt{TU})$ in Section \ref{sec:example}). For example, in Partition $1$, slice $sl_1$ generates a top-up signal and $sl_2$ generates a ramp-down signal or vice-versa. Let us consider the first case, similar arguments follow in the other case too.  There can be three sub-cases in each such case, one giving rise to a top-up-partition action, another to a ramp-down-partition action and the last one demanding neither of the two.  

\textbf{Sub-case 1:} If the requirement of PRBs for the slice opting for top-up (slice $sl_1$ in this example, say), is larger in comparison to the amount of PRBs to be de-allocated from the slice requiring ramp-down (slice $sl_2$), i.e., $\hat{W}_1 > \hat{W}_2$, then the net result is a top-up-partition. This is because slice $sl_1$ demanding a top-up  first borrows $\hat{W}_2$ many PRBs from  slice $sl_2$ that has generated a ramp-down signal. Since the borrowed amount is not sufficient (as $\hat{W}_1 > \hat{W}_2$), some extra PRBs (i.e., amount, $\hat{W}_1 - \hat{W}_2$) are allocated to the partition, resulting in a top-up-partition action.

\textbf{Sub-case 2:} Similarly, if the slice opting for ramp-down (slice $sl_2$), has a greater quantity of PRBs to be reduced compared to the slice requiring top-up (slice $sl_1$), i.e., $\hat{W}_2 > \hat{W}_1$, then the net result is a ramp-down partition. First, the slice asking for ramp-down ($sl_2$) donates $\hat{W}_2$ PRBs to the slice that has demanded for a top-up ($sl_1$). The extra PRBs (i.e., amount $\hat{W}_2 - \hat{W}_1$) that could not be donated, are de-allocated from the partition, causing its ramp-down. 

\textbf{Sub-case 3:} If the number of PRBs to be allocated to one slice is equal to the number to be de-allocated from another slice, i.e., $\hat{W}_1 = \hat{W}_2$, then with intra-partition adjustments within the slices, there is no need to proceed for a top-up-partition or a ramp-down-partition action. 

In all three cases, the PRB-shares (sl-shr$_{i,j}$) and residual PRB counts (sl-resi$_{i,j}$) of the respective slices are updated (i.e., increased for top-up and decreased for ramp-down). The constraint is given as follows.

{\small
{\small \boldmath$L_{2,6}$:} {\it {\large $\land_{k = 1}^{2}$  $\land_{j = 1}^{T}$  } 
 {\large $\land_{i,i' \in \pi_k, \ i \neq i'}$} 
 
 {\huge[} {\large (} sl-top$_{i,j}$ $=$ $\mathbb{T}$  $\land$  sl-ramp$_{i',j}$ $=$ $\mathbb{T}$  {\large )}  
{\large $\Rightarrow$} {\Large[} {\Large\{} {\large(} $\hat{W}_i > \hat{W}_{i'}$ \ $\Rightarrow$ \
 pt-shr$_{k,j}$ $=$
 pt-shr$_{k,j{\text -}1}$ $+$ ( $\hat{W}_i - \hat{W}_{i'}$ ) {\large)} 

 \hspace{7mm} {\Large $\lor$}  {\large(} $\hat{W}_i < \hat{W}_{i'}$ \ $\Rightarrow$ \
 \noindent
 pt-shr$_{k,j}$ $=$ pt-shr$_{k,j{\text -}1}$ $-$ ( $\hat{W}_{i'}$ $-$ $\hat{W}_{i}$ ) {\large )} 
 {\Large $\lor$} {\large(} $\hat{W}_i$ $=$ $\hat{W}_{i'}$  \ $\Rightarrow$ \
 pt-shr$_{k,j}$ $=$ pt-shr$_{k,j{\text -}1}$ {\large)} {\Large \}} \  {\Large $\land$} 

   {\large(} sl-shr$_{i,j}$ $ =$ sl-shr$_{i,j{\text -}1}+\hat{W}_i$ $\land$  
  sl-resi$_{i,j}$  $=$ sl-resi$_{i,j{\text -}1}+\hat{W}_i$  {\large)}   
{\Large $\land$} {\large(}sl-shr$_{i',j}=$ sl-shr$_{i',j{\text -}1}-\hat{W}_{i'}$  $\land$ 
 sl-resi$_{i',j}=$ sl-resi$_{i',j{\text -}1}-\hat{W}_{i'}${\large)} {\Large]} {\huge]}}}

\subsection{Layer 3: Modeling Central Monitoring Agent}
\label{subsec:modeling central agent}
The central monitoring agent in Layer $3$ administers the entire system --- the slices, the partitions and their monitoring agents, as well as the residual partition. 
It plays a vital role in uniformly allocating users to slices and manages the residual partition, making necessary system adjustments if overuse is detected.
Next, we describe the detailed modeling of the central monitoring agent.

\paragraph{{\bf Adjustment of PRB-share of the residual partition:}} As mentioned in Section \ref{sec:RAN model}, the  PRBs that are not allocated to the slices form the residual partition. Hence, whenever a top-up-partition or a ramp-down-partition action occurs, there will be a de-allocation or allocation of PRBs from the residual partition accordingly. There could be another interesting case of an {\it inter-partition adjustment}. For example, Partition 1 has opted for $m$ extra PRBs and Partition 2 wants to free $n$ PRBs, where $n > m$. Thus, first $m$ PRBs are directly given to Partition 1 from Partition 2. The remaining $(n-m)$ will be ramped down from Partition 2 and added to the residual partition (as discussed in Section \ref{sec:example}). There can be several such cases based on the type of actions the individual partitions require; all such cases are handled by the central monitoring agent in Layer 3.

\paragraph{{\bf Uniform allocation of users:}} We have described how the user-count is updated in constraint $L_{1,1}$ in Layer 1 (ref. Section \ref{subsec:constraints layer 1}). Since there are two slices Pre1 and Pre2  for eMBB Premium in our running example, when any user of this service type joins the network, it has to be assigned either to Pre1 or to Pre2. However, random assigning may create clustering of users to only one slice, causing one partition to be overused and another to be underused. For service types having provisions in both partitions (in this case only Premium type users \footnote{Only for the eMBB Premium service type, there are two slices Pre1 and Pre2 in Partitions $1$ and $2$ respectively, each of the other service types have a single slice; Norm1 for eMBB Normal and FWA1 for FWA.}), we design a strategy that analyzes the uniform allocation of users within the partitions and maintains a fair distribution. 

For this, we consider the situation of how far a slice is from its top-up scenario, i.e., a user is assigned to a partition where the corresponding slice (based on the user's service type)
has sufficiently many PRBs before opting for a top-up. As every `$sl{\text -}m$' users use up one PRB in slice $sl$,
thus the PRB-usage is directly proportional to the number of users.

{\it \textbf{Inference 6:}} A slice with a lower number of users has a lower PRB-usage, thus, is farther from a top-up-slice action. 

\noindent
Hence, a user is assigned to a slice with the minimum user-count. There are other service types which are permitted to only one partition, e.g., eMBB Normal and FWA in our example. Such users are directly assigned to their corresponding slices by the central monitoring agent. 

In both these cases, when the residual partition falls below $50\%$ of the total PRBs, allocating additional PRBs to slices is no more feasible, thus preventing new users from acquiring necessary resources. This condition blocks new users from accessing the network.

{\it \textbf{Inference 7:}} The concept of the time-window ($sl{\text -}t{\text -}win$) and determination of the maximum usage, $\lceil \frac{sl{\text -}t{\text -}win}{sl{\text -}m} \rceil$, minimizes the unused PRBs with the help of the partitions' monitoring agents in Layer 2. Moreover, the uniform user allocation in Layer $3$ proportionately assigns users to slices. This prevents overuse of one partition and an increase in the unused resources in another, thereby guaranteeing PRB-optimality.

\subsection{Constraints For Layer 3}
\label{subsec:constraints Layer 3}
Based on the discussion above, we now formally write down the constraints for the central monitoring agent in Layer $3$ for our running example. The new variables introduced in this layer are listed in Table \ref{tab:symbols_layer3}. 
We first describe the residual partition's PRB-share ($rp{\text -}shr_{j}$) update at the j-th timestep, followed by a discussion of the user allocation technique.  \\

\begin{table}[h]
    \small
    \renewcommand{\arraystretch}{1.2}
    \centering
    \begin{tabular}{|l|l|}
        \hline
        \textbf{Symbols} & \textbf{Meaning} \\ \hline
       $rp{\text -}shr_{j}$  &  PRB-share of the residual partition at $j$-th timestep \\ \hline
        $\rho$ & set of partition indices, $\rho=\{1,2\}$, in our example \\ \hline
        $T_P$ & total available PRBs  \\ \hline
        $rp{\text -}ovr_{j}$ &  {\begin{tabular}[c]{@{}c@{}}{residual partition-overuse flag (Boolean variable), is {\it True}} \\ {at $j$-th timestep if residual partition is overused, else {\it False}} \end{tabular}} \\ \hline
        $ser{\text -}prov_{\mu}$ & provision flag (Boolean variable) for the $\mu$-th service type \\ \hline
        $ser{\text -}e_{\mu,j}$ &  user entry flag for $\mu$-th service type at $j$-th timestep \\ \hline
    \end{tabular}
    \caption{New variables associated with Layer 3}
    \label{tab:symbols_layer3}
\end{table}

\noindent
(1) {\bf No change in residual partition:} There is no requirement to add or reduce PRBs from the residual partition 
(i.e., $rp{\text -}shr_{j}=rp{\text -}shr_{j{\text -}1}$) in two cases:

i) Partitions $1$ and $2$ (indexed with $k,k'$) do not opt for top-up or ramp-down-partition action (i.e., case $(\phi_p,\phi_p)$ in Section \ref{sec:example}).
Thus, their PRB-shares remain the same as that in the previous timestep:
i.e., \ $ \land_{k \in \rho} \  (pt{\text -}shr_{k,j} = pt{\text -}shr_{k,j{\text -}1}$),

ii) even if top-up and ramp-down-partition actions occur, the net change is zero because of the inter-partition adjustments, i.e., the number of PRBs de-allocated from one partition ($k$) after a ramp-down is equal to the number of PRBs allocated to the other partition ($k'$) after a top-up, or vice-versa (corresponding to the cases $(\texttt{A},\texttt{D})$ and $(\texttt{D},\texttt{A})$ in Section \ref{sec:example}). This is expressed as: \ \  $\lor_{k, k' \in \rho, \ k \neq k'} \ (pt{\text -}shr_{k,j{\text -}1}$ $-$ $pt{\text -}shr_{k,j}$ = $pt{\text -}shr_{k',j}$ $-$ $pt{\text -}shr_{k',j{\text -}1}$).

The entire constraint is  given as follows. 

{\small
{\small \boldmath$L_{3,1}$:}  {\it {\large $\land_{j = 1}^{T}$}
{\Large [} {\Large \{} {\Large $\land_{k \in \rho}$} {\large (} \ pt-shr$_{k,j}$ $=$ pt-shr$_{k,j{\text -}1}$ 
{\large )}  {\large $\lor$}
{\Large $\lor_{k, k' \in \rho, \ k \neq k'}$} \ {\large (} pt-shr$_{k,j{\text -}1}$ $-$ pt-shr$_{k,j}$ 
 $=$ pt-shr$_{k',j}$ $-$ pt-shr$_{k',j{\text -}1}$ {\large )} {\Large \}} 

\hspace{20mm} 
{\large $\Rightarrow$} rp-shr$_{j}=$rp-shr$_{j{\text -}1}$ {\Large ]} } }\\

\noindent
(2) {\bf Reduction of PRB-share in residual partition:} At least one of the two partitions demands a top-up and the top-up-partition dominates over the ramp-down-partition actions (if occurred). This happens because the neutralization between the partitions' PRB-shares is not applicable here like the previous case. As a result, PRBs are borrowed from the residual partition. 

There are cases where the partitions require only top-up and at first we discuss three such cases. For example, at the $j$-th timestep, any of these three cases can occur: i) case $(\texttt{A},\phi_p)$, i.e., a top-up-partition action in Partition $1$ and no ramp-down/top-up-partition action in Partition $2$  (i.e., $pt{\text -}shr_{1,j} > pt{\text -}shr_{1,j{\text -}1} \land pt{\text -}shr_{2,j} = pt{\text -}shr_{2,j{\text -}1}$), or ii) vice-versa (case $(\phi_p, \texttt{A})$), or iii) case $(\texttt{A},\texttt{A})$ or top-up-partition actions in both the partitions (i.e., $pt{\text -}shr_{1,j} > pt{\text -}shr_{1,j{\text -}1} \land pt{\text -}shr_{2,j} > pt{\text -}shr_{2,j{\text -}1}$). The residual partition's PRB-share is accordingly updated in the respective cases as follows: 

i) $rp{\text -}shr_{j} = rp{\text -}shr_{j{\text -}1} - (pt{\text -}shr_{1,j} - pt{\text -}shr_{1,j{\text -}1})$ or, 

ii) $rp{\text -}shr_{j} = rp{\text -}shr_{j{\text -}1} - (pt{\text -}shr_{2,j} - pt{\text -}shr_{2,j{\text -}1})$ 
or, 

 iii)  $rp{\text -}shr_{j} = rp{\text -}shr_{j{\text -}1} - \ [ \sum_{k \in \rho}(pt{\text -}shr_{k,j} - pt{\text -}shr_{k,j{\text -}1})$] \ ($\rho=\{1,2\}$ as in Table \ref{tab:symbols_layer3}). 

\noindent
Now, at the $j$-th timestep, let there is a top-up-partition action in Partition $1$ and a ramp-down-partition action in Partition $2$ (corresponding to either of the cases $(\texttt{A},\texttt{D})$ and $(\texttt{D},\texttt{A})$ in Section \ref{sec:example}). Also, let the number of PRBs to be allocated to Partition $1$ after a top-up is greater than the number of PRBs to be de-allocated from Partition $2$ after a ramp-down (i.e., $pt{\text -}shr_{1,j} - pt{\text -}shr_{1,j{\text -}1} > pt{\text -}shr_{2,j{\text -}1} - pt{\text -}shr_{2,j}$). In such a case, at first $(pt{\text -}shr_{2,j{\text -}1} - pt{\text -}shr_{2,j})$ many PRBs are ramped down from Partition $2$ and allocated to Partition $1$. The number of PRBs which could not be de-allocated from Partition $2$, i.e., {\it net} = ($(pt{\text -}shr_{1,j} - pt{\text -}shr_{1,j{\text -}1}) - (pt{\text -}shr_{2,j{\text -}1} - pt{\text -}shr_{2,j})$), are borrowed from the residual partition, thereby, reducing its PRB-share, i.e., $rp{\text -}shr_{j} = rp{\text -}shr_{j{\text -}1} - ((pt{\text -}shr_{1,j} - pt{\text -}shr_{1,j{\text -}1}) - (pt{\text -}shr_{2,j{\text -}1} - pt{\text -}shr_{2,j}))$. 
Constraint $L_{3,2}$ is given as follows.

{\small
{\small \boldmath$L_{3,2}$:}  {\it {\large  $\land_{j = 1}^{T}$ }  {\huge [} \ {\Large (}  {\large $\land_{k \in \rho}$} {\large (} pt-shr$_{k,j} >$ pt-shr$_{k,j{\text -}1}$  {\large )} 
$\Rightarrow$ rp-shr$_{j}$ $=$  rp-shr$_{j{\text -}1}$ $-$ $\sum_{k \in \rho}$ (pt-shr$_{k,j}$  $-$ pt-shr$_{k,j{\text -}1}$) {\Large )}  \ {\Large $\lor$}

{\large $\land_{k,k' \in \rho, \ k \neq k'}$} {\Large (} \ {\large (} pt-shr$_{k,j} >$ pt-shr$_{k,j{\text -}1}$  $\land$  pt-shr$_{k',j}$ $=$ pt-shr$_{k',j{\text -}1}$ {\large )} 
$\Rightarrow$ rp-shr$_{j}$ $=$  rp-shr$_{j{\text -}1}$ $-$ (pt-shr$_{k,j}$  $-$ pt-shr$_{k,j{\text -}1}$) \ {\Large )} {\Large $\lor$}

{\large $\land_{k,k' \in \rho, \ k \neq k'}$} {\Large (} \ {\large (}  pt-shr$_{k,j}$ $>$  pt-shr$_{k,j{\text -}1}$ \ $\land$ \ pt-shr$_{k',j}$ $<$ pt-shr$_{k',j{\text -}1}$   
$\land$ \ pt-shr$_{k,j}$ $-$ pt-shr$_{k,j{\text -}1}$ $>$ pt-shr$_{k',j{\text -}1}$ $-$ pt-shr$_{k',j}$ {\large )}  

\hspace{14mm}
{\small $\Rightarrow$} 
{\large (} [ net $=$ ( pt-shr$_{k,j}$  $-$  pt-shr$_{k,j{\text -}1}$ ) $-$ ( pt-shr$_{k',j{\text -}1}$ $-$ pt-shr$_{k',j}$ ) ]
$\land$  rp-shr$_{j}$ $=$ rp-shr$_{j{\text -}1}$ $-$ net {\large )} \ {\Large )} \ \ {\huge ]}  } }\\

\noindent
(3) {\bf Increment of PRB-share in residual partition:} There is a ramp-down partition action for at least one of the two partitions and here the ramp-down-partition actions dominate over the top-up-partition actions (if occurred). Thus, the PRBs de-allocated from the partitions are re-allocated to the residual partition. The sub-cases (i.e., any one of the cases: $(\phi_p,\texttt{D})$, $(\texttt{D},\phi_p)$, $(\texttt{D},\texttt{D})$,  $(\texttt{A},\texttt{D})$ or $(\texttt{D},\texttt{A})$ occur) and the constraint generation in this case are handled similarly like the previous case. 
The constraint is given as follows.  

{\small
{\small \boldmath$L_{3,3}$:}  {\it {\large  $\land_{j = 1}^{T}$ }{\huge [} \ {\Large (}  {\large $\land_{k \in \rho}$} {\large (} pt-shr$_{k,j{\text -}1} >$ pt-shr$_{k,j}$  {\large )} 
$\Rightarrow$ rp-shr$_{j}$ $=$  rp-shr$_{j{\text -}1}$ $+$ $\sum_{k \in \rho}$ (pt-shr$_{k,j{\text -}1}$  $-$ pt-shr$_{k,j}$) {\Large )} \ {\Large $\lor$}

{\large $\land_{k,k' \in \rho, \ k \neq k'}$} {\Large (} \ {\large (} pt-shr$_{k,j{\text -}1} >$ pt-shr$_{k,j}$  $\land$  pt-shr$_{k',j}$ $=$ pt-shr$_{k',j{\text -}1}$ {\large )} 
$\Rightarrow$ rp-shr$_{j}$ $=$  rp-shr$_{j{\text -}1}$ $+$ (pt-shr$_{k,j{\text -}1}$  $-$ pt-shr$_{k,j}$) \ {\Large )} {\Large $\lor$}

{\large $\land_{k,k' \in \rho, \ k \neq k'}$} {\Large (} \ {\large (}  pt-shr$_{k,j}$ $>$  pt-shr$_{k,j{\text -}1}$ \ $\land$ \ pt-shr$_{k',j}$ $<$ pt-shr$_{k',j{\text -}1}$   
$\land$  pt-shr$_{k,j}$ $-$ pt-shr$_{k,j{\text -}1}$ $<$ pt-shr$_{k',j{\text -}1}$ $-$ pt-shr$_{k',j}$ {\large )}

\hspace{14mm}
{\small $\Rightarrow$} 
{\large (} [ net $=$ ( pt-shr$_{k',j{\text -}1}$  $-$  pt-shr$_{k',j}$ ) $-$ ( pt-shr$_{k,j}$ $-$ pt-shr$_{k,j{\text -}1}$ ) ] 
$\land$  rp-shr$_{j}$ $=$ rp-shr$_{j{\text -}1}$ $+$ net {\large )} \ {\Large )} \ \ {\huge ]} } }\\

\noindent
(4) {\bf Uniform user allocation:} Only Premium users can be assigned to any one of the two partitions, as discussed before. For simplicity in writing the following constraint, we associate a flag {\it prov} with every service {\it ser} (i.e., {\it ser-prov}), having value {\it True} (i.e., {\it ser-prov} $=\mathbb{T}$), if the user has a provision in both partitions or else it is {\it False} ($\mathbb{F}$). Here, in our example, there are three service types in total, {\it Premium, Normal} and {\it FWA}, the last two have access in only one partition (i.e., {\it ser-prov} $=\mathbb{F}$).  

We use $\mu$ and $i_{\mu}$ to denote the service type and its corresponding slice indices respectively. The service types {\it Premium, Normal} and {\it FWA} are numbered as $\mu=1,2$ and $3$. Also, as slices {\it Pre1}, {\it Norm1}, {\it Pre2} and {\it FWA} are numbered $1, 2, 3$ and $4$ (ref. Section \ref{subsec:modeling slice}), hence, for $\mu=1$, $i_{\mu}=\{1,3\}$; $\mu=2,i_{\mu}=\{2\}$ and $\mu=3,i_{\mu}=\{4\}$. If any user of the $\mu$-th service type waits to enter the network at some timestep $j$, the variable {\it ser-e$_{\mu,j}$} becomes {\it True}. Moreover, users are allowed to enter the network only if the residual partition is not overused, i.e., $rp{\text-}ovr_{j}=\mathbb{F}$. 

The constraint for updating the user-entry flag ($sl{\text -}en$) of the slices having access in only one partition, is as follows.

{\small
{\small \boldmath$L_{3,4}$:} {\it {\large $\land_{\mu = 1}^3$ $\land_{j = 1}^{T}$}
{\Large [} {\large (} ser-prov$_{\mu} = \mathbb{F}$ $\land$ rp-ovr$_{j} = \mathbb{F}$ $\land$ 
ser-e$_{\mu,j}= \mathbb{T}$ {\large )} 
{$\Rightarrow$} $\land_{i \in i_{\mu}}$ {\large (} sl-en$_{i, j} = \mathbb{T}$ {\large )} {\Large ]} } }\\

Now, we discuss the uniform user allocation corresponding to service type $\mu=1$ or {\it Premium} service type, i.e., allocating users uniformly between the slices Pre1 ($sl_1$) and Pre2 ($sl_3$). At the $j$-th timestep, if the flag $ser{\text -}e_{1,j}$ is {\it True} ($\mathbb{T}$), indicating that a {\it Premium} type user waits to enter the network, then we need to decide in which of the two slices, $sl_1$ and $sl_3$, we can asign the new user. The user-entry flag ({\it sl-en}) of a slice becomes {\it True} if a new user enters the slice $sl$ (ref. Table \ref{tab:symbols_layer1}), i.e., $sl{\text -}en_{i,j}=\mathbb{T}$, when a user is assigned to the $i$-th slice at the $j$-th timestep.

To determine the slice-index with the minimum user-count, we introduce the function {\it Sl-Ind(.)}, that takes the slice's user-count as input and returns the slice-index.
At the $(j{\text -}1)$-th timestep, if $\beta$ is the slice-index having the minimum user-count between slices $sl_1$ and $sl_3$, then the user is assigned to slice $sl_{\beta}$ (i.e., $sl{\text -}en_{\beta,j}=\mathbb{T}$) and not to the other slice (i.e., $\land_{i \in  i_{\mu} - \beta}$ sl-en$_{i j}$ = $\mathbb{F}$).

{\small
{\small \boldmath$L_{3,5}$:} {\it {\Large $\land_{j = 1}^{T}$} 
{\Large [} rp-ovr$_{j}$ $=$ $\mathbb{F}$ \ $\land$ \ ser-e$_{1,j}$ $=$ $\mathbb{T}$
{\large $\Rightarrow$} 
{\Large (} $\beta$ $=$ Sl-Ind (min (sl-usr$_{1,j{\text -}1}$, sl-usr$_{3,j{\text -}1}$) ) \ $\Rightarrow$  
sl-en$_{\beta, j}$ $=$ $\mathbb{T}$ {\large $\land$} $\underset{i \in  i_{\mu} - \beta}{\land}$ sl-en$_{i j}$ $=$ $\mathbb{F}$ {\Large )} {\Large ]} } } \\

For our running example of $2$ partitions and $2$ slices per partition, the entire formal model of the proposed 3-layered framework, {\it FORSLICE}, is obtained by considering the consolidated constraint given as follows.

\hspace{20mm} {\large \textbf{C}: {\it  $ \ \land_{l_{1} = 1}^{6}$ \ $L_{1,\ l_{1}}$ \ $\land$ \ $\land_{l_{2} = 1}^{6}$ \  $L_{2,\ l_{2}}$  \ $\land$ \  $\land_{l_{3} = 1}^{5}$ \ $L_{3,\ l_{3}}$ }} \\

Here, $L_{1,1}-L_{1,6}$ are the constraints in Layer $1$ (ref. Section \ref{subsec:constraints layer 1}), $L_{2,1}-L_{2,6}$ are the constraints in Layer $2$ (ref. Section \ref{subsec:contraints layer 2}) and $L_{3,1}-L_{3,5}$ are the constraints in Layer $3$ (ref. Section \ref{subsec:constraints Layer 3}). If constraint \textbf{C} is {\it satisfiable}, then we conclude that the PRB-allocation generated simultaneously ensures the system properties, fairness and PRB-optimality while prioritizing the eMBB Premium service type. 

Next, we present the constraint formulation in \textit{FORSLICE} for any generic  $(\mathcal{S}, \mathcal{K}, \mathcal{N})$ configuration of the hierarchical 3-layered network, i.e., a 5G network scenario which is much more holistic in terms of PRB-partitioning and slicing.

\subsection{Generalized Constraints for any Service-Partition-Slice Configuration}
\label{appSec:general_constraints}

Here, we consider any number of service types, partitions and slices and also different combinations of slices per partition, abiding by the 5G-network protocols and design criteria. If there are total $\mathcal{K}$ partitions and $r_{k}$ number of slices in the $k$-th partition, $k = 1$ to $\mathcal{K}$, then the total number of slices is $\sum_{k = 1}^{\mathcal{K}} r_{k} =  \mathcal{N}$. 
The slices correspond to various service types. In our running example discussed previously, a service type can have access in more than one partition. For example, eMBB Premium service type had slices Pre1 and Pre2 in Partitions $1$ and $2$ respectively.  

We assume that there are $\mathcal{S}$ service types in total in the network and $n_{\mu}$ number of slices correspond to the $\mu$-th service type, for $\mu = 1$ to $\mathcal{S}$. Therefore, $\mathcal{N} = \sum_{\mu=1}^{\mathcal{S}} n_{\mu}$. We hence obtain the service-partition-slice configuration as $(\mathcal{S}, \mathcal{K}, \mathcal{N})$.
The generalized version of the constraints discussed in Sections \ref{subsec:constraints layer 1}, \ref{subsec:contraints layer 2} and \ref{subsec:constraints Layer 3} (with configuration $(3, 2, 4)$)  are given below.

\textbf{\textit{Layer 1:}} All the constraints mentioned in Section~\ref{subsec:constraints layer 1} for Layer $1$ remain the same except the outer {\it And}, mentioned as $\land_{i = 1}^{4}$ in each of the six constraints ($L_{1,1}-L_{1,6}$), is now replaced by $\land_{i = 1}^{\mathcal{N}}$. We denote the six modified constraints as \boldmath$\mathcal{L}_{1,c}$, where $c = 1$ to $6$.

\textbf{\textit{Layer 2:}} 
To present the generalized constraints for $L_{2,1}-L_{2,6}$ of Layer $2$ in a compact form, we introduce the symbols $\Lambda_{k, 0,j}$, $\Lambda_{k, 1,j}$ and $\Lambda_{k, 2,j}$, that represent the sets containing the slice indices in the $k$-th partition opting for $\phi_s$: {\it no requirements} (neither {\it top-up-slice} nor {\it ramp-down-slice}), $\texttt{TU}$: {\it top-up-slice}, and $\texttt{RD}$: {\it ramp-down-slice} actions, respectively, at the $j$-th timestep.  

It may be the case that the slices in the $k$-th partition either opt for a top-up-slice or a ramp-down-slice action (ref. constraints $L_{2,2}-L_{2,5}$ in Section \ref{subsec:contraints layer 2}). Also, it may happen that a few of the slices in a partition require a top-up-slice action and a few require a ramp-down-slice action; appropriate intra-partition adjustments are made within the slices in such cases (ref. constraint $L_{2,6}$). The PRB-shares of the partitions are accordingly adjusted (increased or decreased), based on the {\it top-up partition} or {\it ramp-down partition} actions. The maximum possible usage in the $i$-th slice between two successive time-windows, $\hat{W}_{i}=\lceil \frac{sl{\text -}t{\text -}win_{i}}{sl{\text -}m_{i}} \rceil$, is allocated/de-allocated to/from every slice after top-up-slice/ramp-down-slice actions, as discussed in Section \ref{subsec:contraints layer 2}. 

Since, $\Lambda_{k, 1,j}$ and $\Lambda_{k, 2,j}$ are the sets of slice indices requiring top-up and ramp-down in $k$-th partition respectively, hence, the total number of PRBs to be allocated to the slices in the partition after top-up-slice actions is $\eta_1=\sum_{u \in \Lambda_{k, 1,j}} \hat{W}_{u}$. 
Similarly, the total number of PRBs to be de-allocated from the slices after ramp-down-slice actions is, $\eta_2=\sum_{v \in \Lambda_{k, 2,j}} \hat{W}_{v}$. 

If $\eta_1>\eta_2$ or $\eta_1<\eta_2$ (former is the case of top-up-slice actions being dominant over ramp-down-slice actions and the latter indicates ramp-down-slice actions are dominant over top-up-slice actions in the $k$-th partition), then the net amount, i.e., $net=\eta_1-\eta_2$ or $net=\eta_2-\eta_1$ is either allocated or de-allocated from the $k$-th partition; also the PRB-shares of the slices are updated accordingly. Else if $\eta_1=\eta_2$, then with intra-partition adjustments, there is no need to allocate or de-allocate PRBs to/from the $k$-th partition; only the PRB-shares of its slices are updated.
The constraint explaining this is given below.

    {\footnotesize
    {\small \boldmath$\mathcal{L}_{2}$:} {\it {\Large $\land_{k = 1}^{\mathcal{K}}$} {\Large $\land_{j = 1}^{T}$}    
    \ {\huge [} {\Large (} $\eta_{1}=\sum\limits_{u \in \Lambda_{k, 1,j}} \hat{W}_{u}$  $\land$ $\eta_{2}=\sum\limits_{v \in \Lambda_{k, 2,j}}\hat{W}_{v}$ {\Large )}
    {\Large $\land$} {\Large \{}  {\Large (} $\eta_{1}>\eta_{2}$  $\Rightarrow$  (net $=\eta_{1} - \eta_{2}$)   
    $\land$ ( pt-shr$_{k,j}=$  pt-shr$_{k,j{\text -}1}+$ net ) {\Large )}     
    
    \hspace{22mm} 
    {\Large $\lor$} {\Large (} $\eta_{2}  > \eta_{1}$  $\Rightarrow$ ( net $= \eta_{2} - \eta_{1}$ )     
    $\land$ ( pt-shr$_{k,j}=$ pt-shr$_{k,j{\text -}1}$ $-$ net )  {\Large )}   
    {\Large $\lor$} {\Large (} $\eta_{1} = \eta_{2}$   
    $\Rightarrow$ ( pt-shr$_{k,j}=$ pt-shr$_{k,j{\text -}1}$ )  {\Large )} \  {\Large \}} 

     \noindent
     {\Large $\land$} {\Large (} {\large $\land_{x \in \Lambda_{k, 0,j}}$} (sl-resi$_{x,j}=$ sl-resi$_{x,j{\text -}1}$)  $\land$  (sl-shr$_{x,j}=$ sl-shr$_{x,j{\text -}1}$) {\Large )} {\Large $\land$}  
     {\Large (} {\large $\land_{u \in \Lambda_{k, 1,j}}$} (sl-resi$_{u,j}=$ sl-resi$_{u,j{\text -}1}$ $+$ $\hat{W}_{u}$) $\land$     
    (sl-shr$_{u,j}=$ sl-shr$_{u,j{\text -}1}$ $+$ $\hat{W}_{u}$)  {\Large )}     
    
    {\Large $\land$} {\Large (} {\large $\land_{v \in \Lambda_{k, 2,j}}$} (sl-resi$_{v,j}=$ sl-resi$_{v,j{\text -}1}$ $-$ $\hat{W}_{v}$) $\land$    
    (sl-shr$_{v,j}=$ sl-shr$_{v,j{\text -}1}$ $-$ $\hat{W}_{v}$ )  {\Large )} {\huge ]} \\} }

\textbf{\textit{Layer 3:}} First we formulate the constraint for PRB-share updation of the residual partition. Let $\mathcal{P}_{0,j}$, $\mathcal{P}_{1,j}$ and $\mathcal{P}_{2,j}$ be the sets containing the partition indices ($1$ to $\mathcal{K}$) which generate $\phi_p$: {\it no change}, $\texttt{A}$: {\it top-up-partition} and $\texttt{D}$: {\it ramp-down-partition} actions at the $j$-th timestep respectively. Based on the combinations of these actions, the residual partition share is suitably adjusted. 

Hence, $\zeta_1=\sum_{u \in \mathcal{P}_{1,j}} (pt{\text -}shr_{u,j} - pt{\text -}shr_{u,j{\text -}1})$ is the number of PRBs to be allocated to the partitions after top-up-partition actions and $\zeta_2=\sum_{v \in \mathcal{P}_{2,j}} (pt{\text -}shr_{v,j{\text -}1} - pt{\text -}shr_{v,j})$ be the number of PRBs to be de-allocated from the partitions after ramp-down-partition actions. 

If $\zeta_1=\zeta_2$, then with inter-partition adjustments, the residual partition share remains the same as that in the previous timestep. Else, if $\zeta_1>\zeta_2$ (top-up-partition actions are dominant over ramp-down-partition actions) or if $\zeta_2>\zeta_1$ (ramp-down-partition actions are dominant over top-up-partition actions), then the PRB-share of the residual partition accordingly decreases or increases.  The following constraint expresses the above idea.

    {\small
    {\small $\mathcal{L}_{3,1}$:} {\it {\Large $\land_{j = 1}^{T}$}
    {\huge [} {\Large (}$\zeta_{1}=\sum_{u \in \mathcal{P}_{1,j}}$(pt-shr$_{u,j} \ -$ pt-shr$_{u,j{\text -}1}$)  {\large $\land$} $\zeta_{2}=\sum_{v \in \mathcal{P}_{2,j}}$(pt-shr$_{v,j{\text -}1}\ -$ pt-shr$_{v,j}$){\Large )} \ {\Large $\land$}

     \hspace{7mm} 
     {\Large [} \ {\Large (} $\zeta_{1} >\zeta_{2} \Rightarrow$ rp-shr$_{j}=$ rp-shr$_{j{\text -}1}- (\zeta_{1}-\zeta_{2})$ {\Large )}       
    {\small $\lor$} {\Large (} $\zeta_{2} > \zeta_{1} \Rightarrow$ rp-shr$_{j}=$ rp-shr$_{j{\text -}1}+(\zeta_{2}-\zeta_{1})$ {\Large )}
    
    \hspace{20mm} 
    {\small $\lor$} {\Large (} $\zeta_{2}=\zeta_{1} \Rightarrow$  rp-shr$_{j}=$ rp-shr$_{j{\text -}1}$ {\Large )} \ {\Large ]} \ {\huge ]} } }\\
 
Next, we discuss the uniform user allocation constraint. Let $n_{\mu}$ be the number of slices corresponding to the $\mu$-th service type, $\mu \le1 \le \mathcal{S}$, i.e., $n_{\mu}=|i_{\mu}|$. These variables are defined above the constraint $L_{3,5}$ in Section \ref{subsec:constraints Layer 3}. Conditions $n_{\mu}=1$ and $n_{\mu}>1$, indicate that the $\mu$-th service has access in only one, and more than one partition respectively.  The generalized constraint $L_{3,5}$ is as follows.

    {\small
    {\small $\mathcal{L}_{3,2}$:} {\it {\Large $\land_{\mu = 1}^{S}$} {\Large $\land_{j = 1}^{T}$}    
    {\Large [}  {\Large (}rp-ovr$_{j}=\mathbb{F}$ $\land$ ser-e$_{\mu,j}=\mathbb{T}$ {\Large )} {\large $\Rightarrow$}

    \hspace{7mm}
    {\huge (} $\beta=$ Sl-Ind {\large (} $\underset{1 \le x \le n_{\mu}}{min}$ ( sl-usr$_{x,j{\text -}1}$ ) {\large )} \ $\Rightarrow$ \ sl-en$_{\beta,j}=\mathbb{T}$ $\ \land \ $ {\Large $\land$} {\small $_{\substack{x = 1 \\ x \neq \beta}}^{n_{\mu}}$} sl-en$_{x,j}=\mathbb{F}$ {\huge )} {\huge ]} } }\\

The final constraint $\mathcal{C} = \land_{i = 1}^{6} \mathcal{L}_{1, i} \ \land \ \mathcal{L}_{2} \ \land_{i = 1}^{2} \ \mathcal{L}_{3, i}$, \ is provided to the SMT solver. An answer, {\it SAT} or {\it satisfiable}, proves that the PRB-allocation generated through the proposed model {\it FORSLICE} satisfies all the system properties while preserving the service level priorities. 

\subsection{Estimation of the Total Number of Constraints}
\label{appSec:number_constraints}

Here, we obtain an upper bound on the total number of constraints formulated in all  three layers.

\textbf{Layer $1$:}  Layer $1$ comprises of the slices and each slice is modeled with the set of constraints, $\land_{i = 1}^{6} \mathcal{L}_{1, i}$.  
Hence, there are $6 \times T \times \mathcal{N}$ many constraints in Layer $1$ in total, on simulating till the timestep $T$.

\textbf{Layer $2$:} In Layer 2, each of the $r_k$ slices in the $k$-th partition can opt for exactly one of the following three types of actions at the $j$-th timestep: {\it no action} (neither top-up nor ramp-down), {\it top-up-slice}, and {\it ramp-down-slice}. This leads to a total of $3^{r_{k}}$ possible combinations at each timestep, each of which is modeled with a constraint. All such possible cases are mentioned in constraint $\mathcal{L}_{2}$.  
As the same argument holds at each timestep, the total number of constraints in Layer $2$, up to timestep $T$, is $\sum_{k = 1}^{\mathcal{K}} T \times 3^{r_{k}}$.

\textbf{Layer $3$:} 
For the residual partition share management, similar to the monitoring agents in the second layer, the central monitoring agent too needs to adhere to three type of actions --- {\it no action} (PRB-shares in partitions remain same), {\it top-up-partition}, and {\it ramp-down-partition} --- to decide on the PRB-share of the residual partition. Consequently, the central monitoring agent deals with $3^{\mathcal{K}}$ combinations for $\mathcal{K}$ many partitions in total.

Moreover, the uniform user allocation to slices is aided by the determination of the minimum user-count amongst all the slices of the same service type. In order to calculate the minimum value and evaluate the value of $sl{\text -}en_{i,j}$ to {\it True} or {\it False}, $n_{\mu}$ comparisons and $n_{\mu}$ update equations (both checked through constraints) are performed at each timestep, for $n_{\mu}$ many slices corresponding to the $\mu$-th service type.

Following the above arguments, the total number of constraints executed in Layer $3$, up to timestep $T$ is, $T \times (3^{\mathcal{K}} + \sum_{\mu = 1}^{\mathcal{S}} 2 \times n_{\mu})$.

Taking into account the internal sub-constraints, we state that the total number of constraints in all the three layers is of $O(T \times( 6 \mathcal{N} + \sum_{k = 1}^{\mathcal{K}} 3^{r_{k}} + 3^{\mathcal{K}} + \sum_{\mu = 1}^{\mathcal{S}} 2 n_{\mu})) \ = \ O(f(T, \mathcal{S}, \mathcal{K}, \mathcal{N}))$.

This wraps up the entire formal modeling of \textit{FORSLICE} and next we present a set of experiments to justify that the proposed framework efficiently ensures the system properties.

\section{Experimental Evaluation}
\label{sec:results}

In this section, we evaluate the proposed formal framework {\it FORSLICE}, based on the following four research questions.
\begin{compactenum}[Q1.]
    \item  How effectively does \textit{FORSLICE} maintain the system properties, \textit{fairness} and \textit{PRB-optimality}?
    \item How does the PRB-allocation in \textit{FORSLICE} benefit best-effort services?  
    \item How does increasing the number of timesteps affect the execution time of \textit{FORSLICE}?
    \item How does \textit{FORSLICE} compare with the baseline~\cite{convergence} in terms of system properties?
\end{compactenum}

\paragraph{\textbf{Implementation:}} 
To address the above questions, we evaluate {\it FORSLICE} by considering various types of input configurations, specifically by varying the following input parameters.

    (1) Total number of service types ($\mathcal{S}$), \ \ 
    (2) Total number of partitions ($\mathcal{K}$), \ \
    (3) Total number of slices ($\mathcal{N}$), \  \ 
    (4) Number of slices per partition

Different $(\mathcal{S}, \mathcal{K}, \mathcal{N})$ configurations are created by varying the above parameters. For example, in the $(3, 2, 4)$ configuration discussed in Section \ref{sec:example}, we considered $3$ service types with $2$ partitions and $4$ slices; Partitions $1$ and $2$ contains $2$ slices each. Here, we account for three other types of configurations, viz., $(3, 3, 7)$, $(5, 3, 10)$ and $(5, 4, 13)$, where either $\mathcal{S}$ or $\mathcal{K}$ is increased, thereby increasing the total number of slices, $\mathcal{N}$. Note that an increase in the number of service types, $\mathcal{S}$, contributes to a larger number of users in the network. This increases the number of slices per partition (thus the total number of slices $\mathcal{N}$) to cater the users with differentiated quality of service depending on their needs. Moreover, $\mathcal{N}$ automatically increases with an increase in the number of partitions ($\mathcal{K}$), since a single partition has at least one slice.

Consider the $(3, 3, 7)$ configuration. We fix the number of service types ($\mathcal{S}$) to $3$ as in $(3, 2, 4)$, and increase one more partition (from $2$ to $3$) and the total number of slices $\mathcal{N}$ (from $4$ to $7$), to obtain $3$ partitions with $7$ slices. On the other hand for the $(5, 3, 10)$ configuration, keeping the number of partitions ($\mathcal{K}$) fixed, we increase the number of service types $\mathcal{S}$ (from $3$ to $5$) and the number of slices per partition, which in turn increases the total number of slices $\mathcal{N}$ (from $7$ to $10$). Finally, for the $(5, 4, 13)$ configuration, we keep the parameter $\mathcal{S}$ fixed, as in $(5, 3, 10)$, and increase both $\mathcal{K}$ and $\mathcal{N}$ to construct $4$ partitions with $13$ slices. Note that with an increase in the total number of services ($\mathcal{S}$), partitions ($\mathcal{K}$) and slices ($\mathcal{N}$), the number of users per partition also increases. Thus, while evaluating \textit{FORSLICE} by considering multiple ($\mathcal{S}$, $\mathcal{K}$, $\mathcal{N}$) configurations, we simultaneously examine its effectiveness by varying the total user count.

The 5G service types considered in this work are: enhanced mobile broadband (eMBB) --- eMBB Premium (Pre) and eMBB Normal (Norm), fixed wireless access (FWA), ultra-reliable low-latency communications (URLLC) and massive machine-type communications (mMTC). 
We have chosen the standard service types from the 3GPP specification \cite{5GS_3GPP} for our experiments.
Considering up to $13$ slices at maximum for $5$ service types is in accordance with the standard network protocol and design criteria followed for 5G telecommunication networks (Table 5.15.2.2-1 on Page 23 in ~\cite{3GPP_SST} mentions that are 6 standardized service-slice types (SST), whereas, we test up to 13 slices so that it can handle future standard SSTs).

The user distributions vary according to the service types; for example, in usual practice, eMBB services are often characterized by heavy-tailed distributions (such as lognormal), whereas FWA services typically follow a Poisson distribution, etc. Based on such user distributions, we construct the user inputs (arrival and leaving of users) to our formal model. 

We first generate random data following the stipulated distributions and then convert the generated probability density values (according to the probability density function or mass function) into binary flags $(0/1)$, using thresholds. If the probability is above the given threshold for the $i$-th slice at the $j$-th timestep, then we set the user entry flag \textit{sl-en}$_{i,j}=1$, else we set \textit{sl-en}$_{i,j}=0$ (ref. Table \ref{tab:symbols_layer1} for the definition of \textit{sl-en}$_{i,j}$).

With the generated user data corresponding to a particular $(\mathcal{S}, \mathcal{K}, \mathcal{N})$ network configuration and the simulation time horizon given as input, we simulate \textit{FORSLICE} (see the automated workflow in Figure \ref{fig:automated_workflow}) for a finite number of timesteps, to obtain the PRB-allocations of slices and partitions as outputs. 
The workflow provides a fully automated, end to end solution --- processing the inputs to formulate all the three-layer constraints, invoking an SMT solver for verification, and finally generating PRB-allocations if the set of constraints are satisfiable.

From the allocations generated as output, we can compute other parameters, such as, \textit{number of top-up-slice and ramp-down-slice actions} (based on the number of times PRBs have been allocated/de-allocated to/from slices), \textit{residual partition share} (the PRBs not allocated to the slices), etc. These parameters allow us to resolve the research questions stated previously.

\paragraph{\textbf{Experimental Setup:}} We have used the  Z3 SMT solver~\cite{z3} with Python API to formally model \textit{FORSLICE}. The user-data generation, following various probability distribution functions, is implemented  by integrating the `scipy.stats' module in Python. All these experiments are carried out on a 64-bit Windows OS in a 2.10 GHz Intel Core-i5 machine, with 32 GB of RAM. The network simulations are performed by connecting user equipments (UEs) to a single base station (gNB) in the network simulator NS3-5Glena \cite{5G_NS_Glena}.

\subsection{Q1. Checking Fairness and PRB-Optimality}
\label{subsec:results_Q1}
Here, we present two sets of experiments to demonstrate that {\it FORSLICE} ensures the system properties: fairness and PRB-optimality.

\subsubsection{Residual Partition Share}
\label{subsec:results_Q1_residual_partition}

\begin{wrapfigure}{r}{0.48\linewidth}  
	\centering
	\includegraphics[scale=0.3]{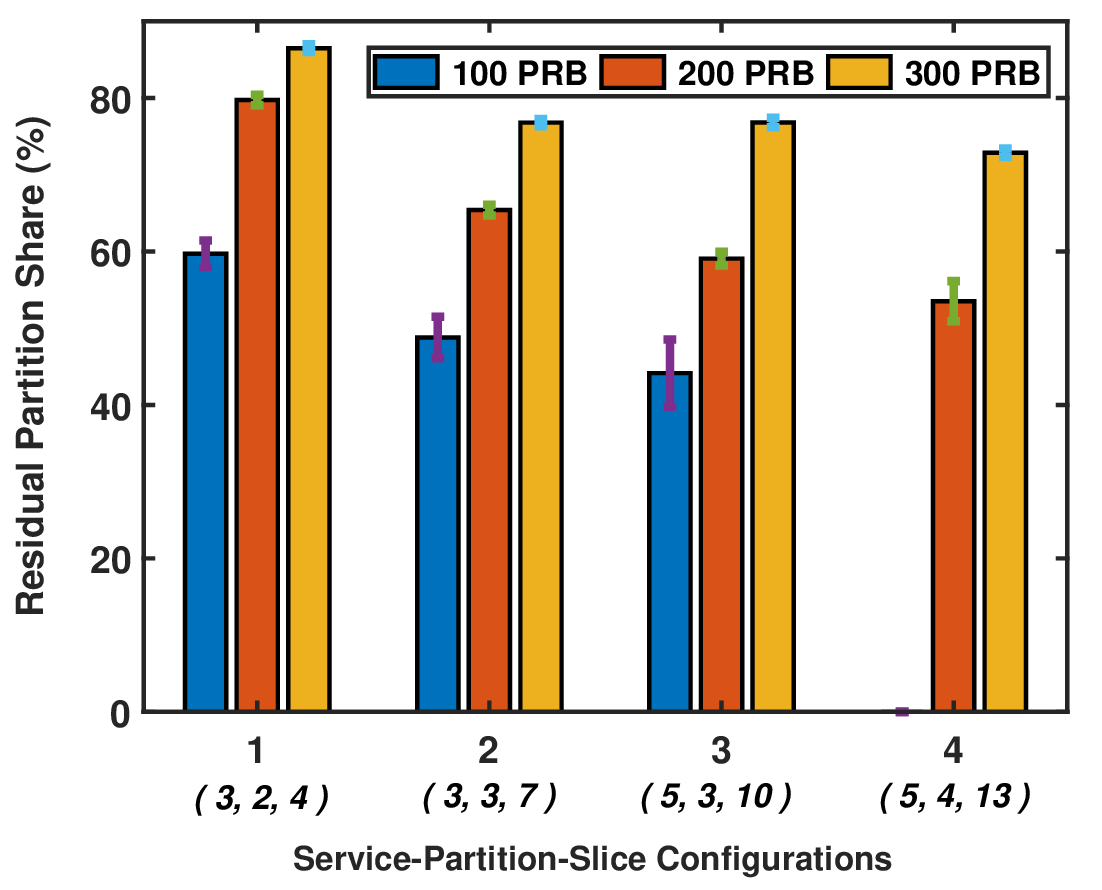}
	\caption{The final residual partition share for different PRB counts in four \boldmath$(\mathcal{S}, \mathcal{K}, \mathcal{N})$ configurations, for 30 timesteps}
	\label{fig:Resi-share}
\end{wrapfigure}

First, we check the fairness by analyzing the PRB-share of the residual partition.
Figure \ref{fig:Resi-share} shows how the final residual partition shares vary with the four $(\mathcal{S}, \mathcal{K}, \mathcal{N})$ configurations and for three different values ($100$, $200$ and $300$) of the total PRB count, when {\it FORSLICE} is simulated for $30$ timesteps, e.g., for $\SI{30}{\minute}$ (with one timestep $h=\SI{1}{\minute}$). Each bar plot in the figure depicts the PRB-share of the residual partition in the final timestep (i.e., the $30$-th timestep) for a particular $(\mathcal{S}, \mathcal{K}, \mathcal{N})$ configuration and a fixed total PRB count. 

To generate the data for each bar plot, we ran the \textit{FORSLICE} simulation $30$ times generating different instances of user data (\textit{sl-en}$_{i,j}$ and \textit{sl-lv}$_{i,j}$) in each trial. Specifically, our evaluation utilized a large volume of user data instances.
The corresponding confidence interval, obtained considering $30$ runs of the simulation for a single case, is also marked in each bar plot.

The $(5, 4, 13)$ configuration exhibits a bit of anomaly, with respect to the total PRB count.
There are $4$ partitions and $13$ slices in total for this configuration, and due to the uniform allocation of users to the partitions by the central monitoring agent (ref. $\mathcal{L}_{3,2}$, generalized constraint for $L_{3,5}$), users are almost equally spread into the four partitions, if they have provisions in more than one partition. Thus, $100$ PRBs, is too low to be allocated to the users among $13$ slices.
Hence, for this configuration, the results are shown only for the cases with total PRB as $200$ or $300$. For the rest of the three configurations, the results for all three cases of $100$, $200$ and $300$ total PRBs are shown in Figure \ref{fig:Resi-share}. 

It is clearly evident from the figure that with an increase in the number of slices and partitions, the PRB utilization is greater (due to the larger number of users assigned to the slices) and therefore the PRB-share of the  residual partition gradually drops. 
Furthermore, for all four configurations and for all total PRB counts, the PRB-share of the residual partition is at least $50\%$ of the total PRBs, which implies that the residual partition is not yet overused. This is specifically achieved due of the design of the following constraints in Layer 3 of our model: i) inter-partition adjustments to manage the residual partition PRB-share (ref. constraint $\mathcal{L}_{3,1}$ in Section \ref{appSec:general_constraints}), and ii) uniform allocation of users within partitions (ref. constraint $\mathcal{L}_{3,2}$ in Section \ref{appSec:general_constraints}). These actions reduce clustering of users in one partition and prevent continuous overuse of the residual partition. 

Therefore, the PRB-share of the residual partition (denoted by the variable \textit{rp-shr}$_j$ at the $j$-th timestep, ref. Table \ref{tab:symbols_layer3}) does not get overused. Consequently, the boolean variable \textit{rp-ovr}$_{j}$ (that becomes \textit{True} in case of a residual partition-overuse scenario, ref. Table \ref{tab:symbols_layer1}) remains \textit{False} ($\mathbb{F}$) at the $30$-th timestep. This implies that a top-up signal (\textit{sl-top}) can be generated in a slice whenever there is PRB-requirement from the user end (ref. constraint $L_{1,4}$). Based on the formal modeling of {\it FORSLICE}, extra PRBs are allocated to slices upon the generation of top-up signals (ref. constraints $L_{2,2}, L_{2,4}$ and $ L_{2,6}$), ensuring fairness to each service type, as per Definition \ref{def:fairness}.

\subsubsection{Top-up-slice and Ramp-down-slice Actions}
\label{subsec:results_Q1_topup_rampdown}

We now demonstrate that {\it FORSLICE} strikes a balance between the total number of top-up-slice and ramp-down-slice actions, to ensure fairness and PRB-optimality simultaneously. Top-up in a slice is required to guarantee fairness to the users of the slice corresponding to a particular service type, whereas ramp-down in a slice is necessary to reduce the extra and unused PRBs from a slice, minimizing the PRB-usage. The proposed model aims to reduce the overall dynamic actions: top-up-slice and ramp-down-slice actions, while ensuring fairness and PRB-optimality. The number of top-up-slice and ramp-down-slice actions in a duration of $30$ timesteps, for a total PRB count of 200, is exemplified in Table \ref{tab:top-up,ramp-down}.

\begin{table}[t]
   \small \sf
   \renewcommand{\arraystretch}{1.25}
    \centering
    \begin{tabular}{|c|c|c|c|}
        \hline
       \textbf{Row No.} & \begin{tabular}[c]{@{}c@{}}  \boldmath$(\mathcal{S}, \mathcal{K}, \mathcal{N})$ \\ \textbf{Configuration}  \end{tabular} &  \begin{tabular}[c]{@{}c@{}} \textbf{\# Top-Up-} \\ \textbf{Slice Actions} \end{tabular} & \begin{tabular}[c]{@{}c@{}} \textbf{\# Ramp-Down} \\ \textbf{Slice Actions} \end{tabular} \\ \hline
        1 & $(3, 2, 4)$ & 6 & 0 \\ \hline
        2 & $(3, 3, 7)$ & 7 & 0 \\ \hline
        3 & $(5, 3, 10)$ & 6 & 3 \\ \hline
        4 & $(5, 4, 13)$ & 10 & 0 \\ \hline
    \end{tabular}
    \caption{{Total number of top-up-slice and ramp-down-slice actions for 200 PRBs}}
    \label{tab:top-up,ramp-down}
\end{table}

We observe that when both partitions and slices are increased, the number of top-up-slice actions also increases to provide the users with the desired amount of PRBs ($\lceil \frac{sl{\text -}t{\text -}win}{sl{\text -}m} \rceil$ many PRBs during one time interval) and thereby guarantee fairness. However, interestingly the rate of increase is not that high. For example, corresponding to the $(3, 2, 4)$ configuration, there are $6$ top-up-slice actions in $4$ slices after simulating for $30$ timesteps; whereas there are just $7$ top-up-slice actions in a larger number of slices ($7$ slices) for the $(3, 3, 7)$ configuration (Rows $1{\text -}2$). 

The primary reason behind this is as follows. A top-up-slice action allocates $\lceil \frac{sl{\text -}t{\text -}win}{sl{\text -}m} \rceil$ many extra PRBs (the maximum usage amount) to a slice $sl$ after a top-up, such that there are sufficiently many PRBs till the next time window (according to Inference $5$). Hence, within a time-interval, there is no need for a top-up in a slice; the action is commenced only if it is truly essential to allocate extra PRBs further (i.e., when the residual PRB in a slice, sl-resi$_{i,j}$, is less than the amount, $\lceil \frac{sl{\text -}t{\text -}win}{sl{\text -}m} \rceil$, as per constraint $L_{1,4}$).

Furthermore, with an increase in both the number of partitions and slices, the users are uniformly allocated within the partitions. The PRBs are thus uniformly utilized by the users and  do not remain unused in a particular slice due to overuse in another. Therefore, it is not required to evoke ramp-down-slice actions often to de-allocate the excess unused PRBs.

However, keeping the number of partitions fixed, if only the number of slices is increased, then the clustering of users becomes larger in one partition, resulting in its overuse and a low PRB usage in another. This increases the chances of a ramp-down-slice action in the slice with a larger number of unused PRBs. The unused PRBs can be then ramped down and offered to another slice which is in requirement of extra PRBs (ref. constraint $L_{2,6}$ in Layer $2$). Such a scenario is reported in Rows $2{\text -}3$; where for the same set-up of $3$ partitions, the number of slices increases from $7$ to $10$, thereby escalating the ramp-down slice actions in the latter case. Nevertheless, this simultaneously minimizes the unused PRBs in slices, ensuring PRB-optimality (according to Inference $5$ and $7$). 

In summary, \textit{FORSLICE} preserves the system properties, fairness and PRB-optimality, however, it tries to evoke the top-up-slice and ramp-down-slice actions as less as possible to avoid unnecessary allocations and de-allocations at the runtime.

\subsection{Q2. Benefiting Best-Effort Services}
\label{subsec:results_Q2}
The PRB-allocation approach in this work also aims to cater the best-effort services (services having no specific QoS requirements like eMBB, FWA), apart from ensuring fairness and PRB-optimality while prioritizing the eMBB Premium service type.
The following two strategies in \textit{FORSLICE} specifically make provision for the best-effort services that access the residual partition. 

i) A top-up signal can be generated in a slice only when the residual partition is not overused, i.e., when it has more than $50\%$ of the total PRBs (ref. constraint $L_{1,4}$ in Layer $1$). If overused, we avoid a top-up-slice action, unless PRBs are re-allocated to the residual partition after some time. 

ii) On having excess unused PRBs in a slice for a long duration, the unused amount is ramped-down from the partition (corresponding to that slice) and re-allocated to the residual partition (ref. constraint $L_{1,5}$ in Layer $1$ and constraint $L_{3,3}$ in Layer $3$). 

iii) The inter-partition adjustments in Layer $3$ often lead to scenarios in which PRBs need not to be de-allocated from the residual partition after top-up-slice actions (ref. constraint $L_{3,1}$ in Layer $3$).

We now discuss experimental results illustrating how best-effort services gain from the PRB-share management of the residual partition, which is particularly achieved due to the aforementioned strategies.

We observe in Figure \ref{fig:Resi-share} that the final residual partition share is much greater than $50\%$ of the total number of PRBs, in all cases. It is illustrated in Section \ref{subsec:results_Q1_residual_partition} that such  scenarios ensure fairness to every service type, by providing the facility for a top-up-slice action in times of PRB-requirement. Thus, the proposed model \textit{FORSLICE} allocates PRB in such a manner, which apart from guaranteeing fairness, also allows for a sufficient amount of PRBs ($50\%$ of the total amount) to be utilized by multiple best-effort services. 

Additionally, the illustration and the results provided in Section \ref{subsec:results_Q1_topup_rampdown} indicate that the rate of increase in the number of top-up-slice actions is quite low, even with increasing slices and partitions. Moreover, ramp-down-slice actions are invoked to reduce unused PRBs when there is an overuse in one partition and PRBs remain relatively unutilized in another. Both these factors prevent the residual partition from getting overused, highlighting that the PRB-allocations/de-allocations in \textit{FORSLICE} ensure PRB-optimality and simultaneously accounts for the residual partition to have 50\% of the total PRBs.

\subsection{Q3. Execution Time}
\label{subsec:results_Q3}
Here, we explain how the \textit{execution time} of \textit{FORSLICE} varies with the number of \textit{timesteps}. 
For this experiment, we consider three input combinations: $30$, $50$ and $70$ timesteps (e.g., $\SI{30}{\minute}$, $\SI{50}{\minute}$ and $\SI{70}{\minute}$), respectively, to observe the rate of increase in the execution time with higher values of timesteps. For all four $(\mathcal{S},\mathcal{K},\mathcal{N})$ configurations and for different values of timesteps, the respective minimum and average verification times (averaged over $3$ runs) are shown in Table \ref{tab:timesteps}. 

\begin{table}[h]
  \small \sf
  \renewcommand{\arraystretch}{1.25}
  \centering
   \begin{tabular}{|l|ccc|ccc|}
        \hline
        \boldmath$(\mathcal{S},\mathcal{K},\mathcal{N})$  & \multicolumn{3}{c|}{\textbf{( 3, 2, 4 )}} & \multicolumn{3}{c|}{\textbf{( 3, 3, 7 )}} \\ \hline
        \rowcol\begin{tabular}[l]{@{}l@{}} \textbf{Time}\textbf{steps}  \end{tabular}   & \multicolumn{1}{c|}{30}   & \multicolumn{1}{c|}{50}  & 70  & \multicolumn{1}{c|}{30}  & \multicolumn{1}{c|}{50}  & 70      \\ \hline
        \rowcol \begin{tabular}[l]{@{}l@{}} \textbf{Min} \textbf{Time (s)}  \end{tabular} & \multicolumn{1}{c|}{3.50}  & \multicolumn{1}{c|}{32.97} & 157.31 & \multicolumn{1}{c|}{23.67} & \multicolumn{1}{c|}{41.14}  & 366.65 \\ \hline
        \rowcol \begin{tabular}[l]{@{}l@{}} \textbf{Avg} \textbf{Time (s)}  \end{tabular} & \multicolumn{1}{c|}{12.49} & \multicolumn{1}{c|}{94.49} & 439.20 & \multicolumn{1}{c|}{32.66} & \multicolumn{1}{c|}{100.57} & 417.99 \\ \hline \hline
        \boldmath$(\mathcal{S},\mathcal{K},\mathcal{N})$    & \multicolumn{3}{c|}{\textbf{( 5, 3, 10)}}  & \multicolumn{3}{c|}{\textbf{(5, 4, 13)}}   \\ \hline
        \rowcol \begin{tabular}[l]{@{}l@{}} \textbf{Time}\textbf{steps}  \end{tabular} & \multicolumn{1}{c|}{30}     & \multicolumn{1}{c|}{50}     & 70      & \multicolumn{1}{c|}{30}     & \multicolumn{1}{c|}{50}      & 70      \\ \hline
        \rowcol \begin{tabular}[l]{@{}l@{}} \textbf{Min} \textbf{Time (s)}  \end{tabular}  & \multicolumn{1}{c|}{74.688}  & \multicolumn{1}{c|}{183.05} & 457.15 & \multicolumn{1}{c|}{116.61} & \multicolumn{1}{c|}{977.15}  & 4145.75 \\ \hline
        \rowcol \begin{tabular}[l]{@{}l@{}} \textbf{Avg} \textbf{Time (s)}  \end{tabular}  & \multicolumn{1}{c|}{133.63} & \multicolumn{1}{c|}{319.88} & 1195.17 & \multicolumn{1}{c|}{216.90} & \multicolumn{1}{c|}{1439.53} & 5037.29 \\ \hline
    \end{tabular}
\caption{Execution time with varying timesteps} 
\vspace{-1mm}
\label{tab:timesteps}
\end{table}

Through our experiments, we observe that the execution time increases with increasing timesteps, as well as with an increase in the number of slices and partitions. For a smaller number of slices, e.g., for $4$ slices, the execution time taken is mostly within $\SI{12}{\second}$-$\SI{7}{\minute}$, whereas, for $7$ slices, it is approximately around $\SI{7}{\minute}$, up to $70$ timesteps. For $10$ slices, the  execution time is less than $\SI{5.5}{\minute}$ up to $50$ timesteps; it increases to $\SI{20}{\minute}$ in average for higher timesteps. Finally, for $4$ partitions and $13$ slices, the time varies between $\SI{3.5}{\minute}$-$\SI{24}{\minute}$ up to $50$ timesteps, and increases quite a bit, nearly $\SI{1.4}{\hour}$, for 70 timesteps. Therefore, in almost all cases, {\it FORSLICE} can be executed within a reasonable time frame; e.g., within half an hour.  

We consider the user arrival data following the user distributions for specific service types, compliant with the 5G telecommunication network protocols. Also, the number of slices is considered following real practice, $8-13$ in general. Since our modeling adheres to such standard design criteria and yields reasonable execution times for PRB-allocation over a given simulation time range, we are confident that \textit{FORSLICE} is practically viable for deployment in real applications.

\subsection{Q4. Comparison with Baseline}
\label{subsec:results_Q4}

To our knowledge, the only existing work that considers similar PRB-partitioning and RAN-slicing scenarios, accounting for fairness and prioritizing eMBB Premium service, is \cite{convergence}. Their proposed method \textit{`Convergence'} employs an AI planning agent to solve the PRB-allocation problem. However, the method is limited to the PRB-allocation solution only for a single case study: $3$ service types, $2$ partitions and $2$ slices per partition (Pre1, Norm1 in Partition 1 and Pre2, FWA1 in Partition 2).

\begin{wrapfigure}{r}{0.48\linewidth}
  \centering
    \includegraphics[scale=0.3]{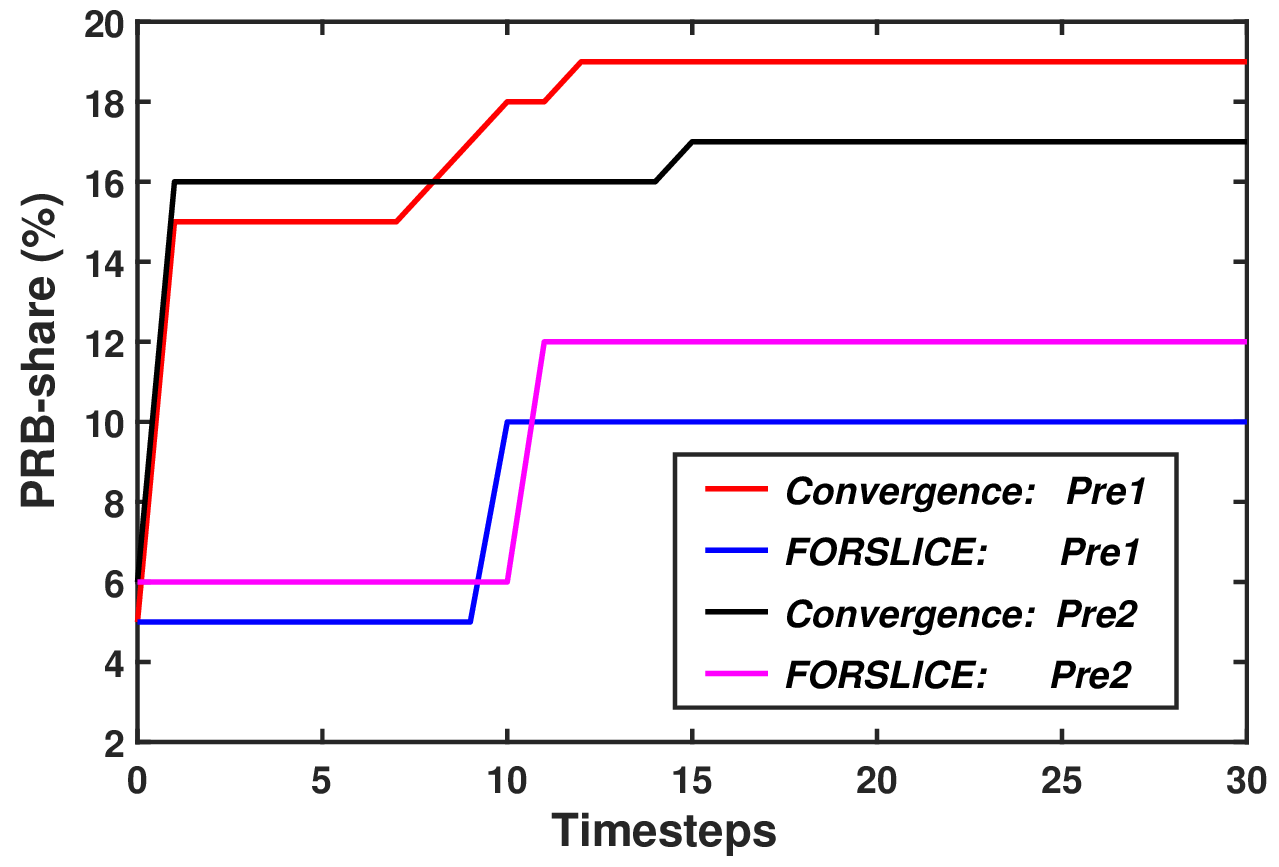}
  \caption{\small Comparing $\%$ of PRB-share in eMBB Premium slices with \textit{Convergence}}
  \vspace{-1mm}
  \label{fig:comparison}
\end{wrapfigure}

In contrast, the proposed formal framework, {\it FORSLICE}, is generic and automated, and successfully computes PRB-allocations for any configuration of services, partitions and slices. Moreover, our work addresses residual-partition share management in order to serve the best-effort services, which is not accounted for in \cite{convergence}.

To demonstrate the efficiency of {\it FORSLICE}, we compare the \textit{PRB-share percentage} in eMBB Premium slices, with the AI-planning based solution method, \textit{Convergence}. 
Since both the methods prioritize the eMBB Premium service type, we consider the metric, PRB-share in eMBB Premium slices, to highlight how {\it FORSLICE} simultaneously ensures fairness and PRB-optimality, with a priority to the eMBB Premium service type, in contrast to \textit{Convergence}.

For this experiment, we consider the $(3, 2, 4)$ configuration, i.e., $2$ partitions and $2$ slices per partition (Pre1, Norm1 in Partition 1 and Pre2, FWA1 in Partition 2), since this is the only configuration considered in ~\cite{convergence}. Finally, we simulate both models for $30$ timesteps.

Starting with the same initial PRB-share, it is clear from Figure \ref{fig:comparison} that at all timesteps, the percentage of the PRB-allocations to any of the slices, Pre1 or Pre2, in \textit{Convergence}, is greater than that in {\it FORSLICE}. Although this prioritizes eMBB Premium serivce type and guarantees fairness in \cite{convergence}, PRBs are over-provisioned to the slices in their case (at least by $44.45\%$ as compared to ours). Rather, we ensure fairness with a much lower percentage of allocated PRBs, as we minimize the allocated but unused PRBs in the slices. 

This comparison thus highlights the PRB-optimality in {\it FORSLICE} (Inferences 5, 7). Moreover, the residual partition PRB-share is larger in our case (as allocations to slices are lesser), thereby benefiting the best-effort services.

\section{A Case Study: Network Simulation}
\label{sec:network_simulation}

Here, we consider a practical example of Ericsson's PRB-partitioning in Saltlake Kolkata, to discuss the network performance achieved in \textit{FORSLICE}. There is a diverse demand for services in that region, like eMBB (demanding a large share of PRBs during peak evening hours for high throughput, especially in the office areas), FWA (having PRB requirements to ensure a minimum throughput), etc. To provide guaranteed SLAs and traffic isolation for different services running over the same physical infrastructure, efficient PRB-allocations is a necessity.

\begin{figure}[h]
  \centering
  \begin{subfigure}{.5\textwidth}
    \centering
    \includegraphics[scale=0.23]{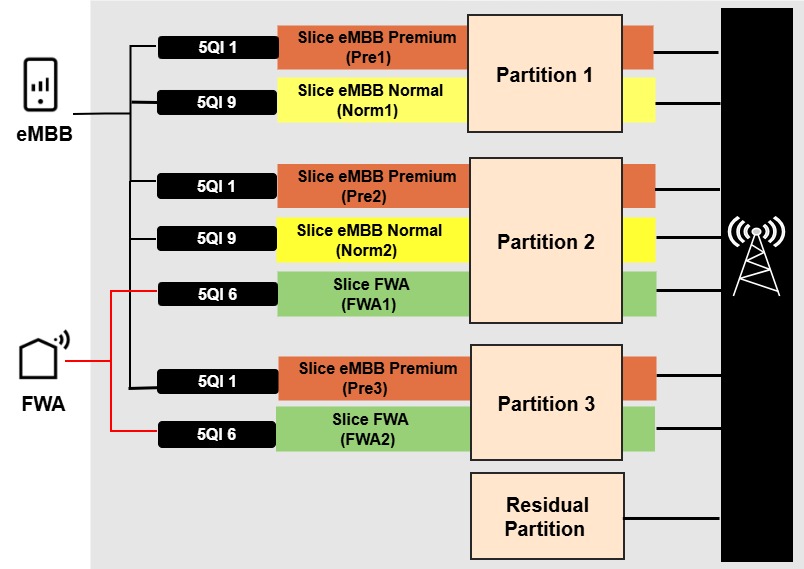}
    \caption{RAN slicing scenario}
    \label{fig:network_simulation_slice_partition}
  \end{subfigure}%
  \begin{subfigure}{.5\textwidth}
    \centering
    \includegraphics[scale=0.195]{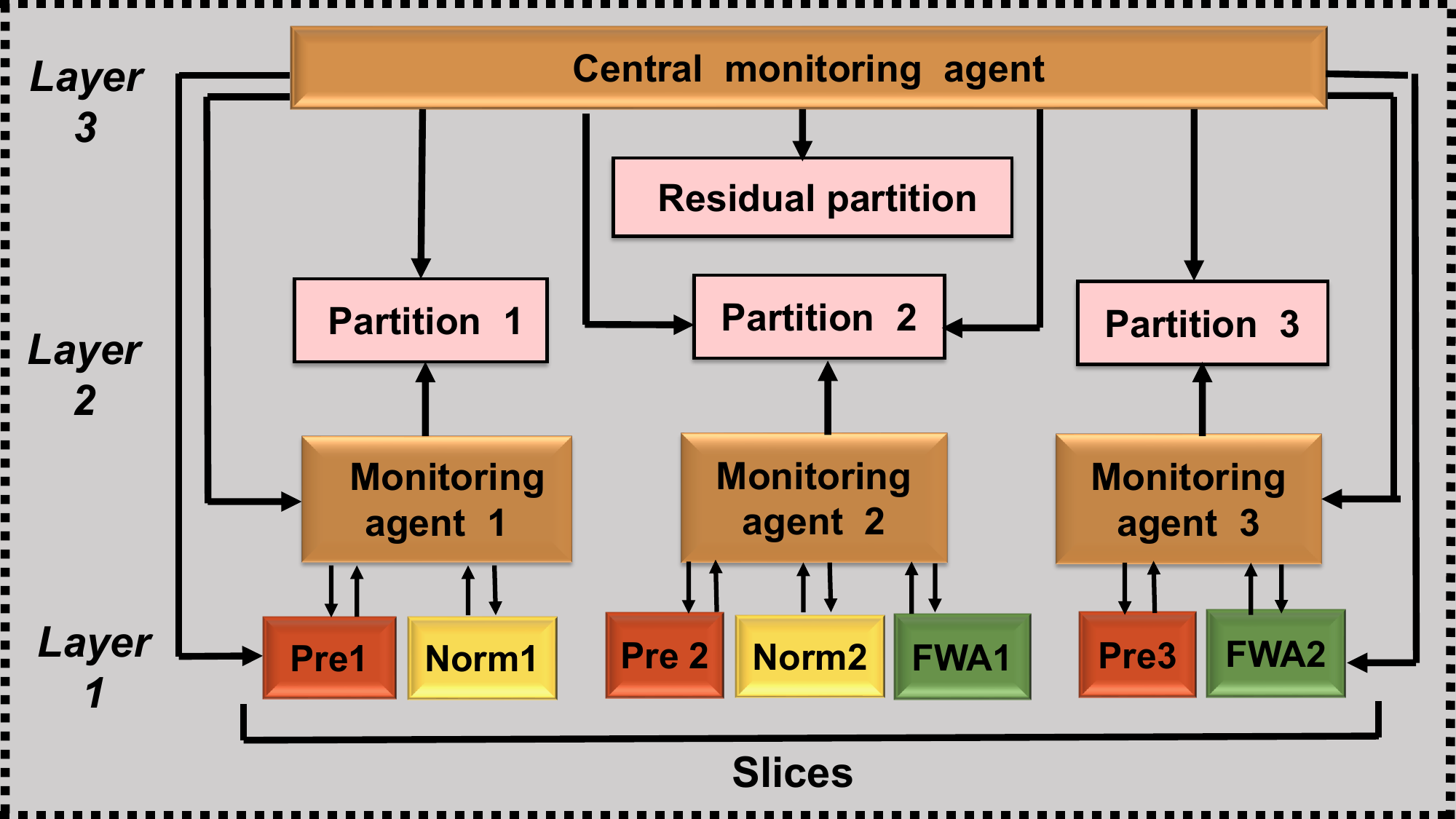}
    \caption{3-layered \textit{FORSLICE} framework}
    \label{fig:network_simulation_FORSLICE}
  \end{subfigure}
  \caption{Network configuration $(3, 3, 7)$: service types eMBB Premium, Normal and fixed wireless traffic; slices Pre1 and Norm1 in Partition 1;  Pre2, Norm2 and FWA1 in Partition 2;  Pre3 and FWA2 in Partition 3}
  \label{fig:network_simulation_RAN}
\end{figure}

Based on the user categories and resource demands, we have considered 3 partitions with the slice types as eMBB Premium, eMBB Normal and FWA. Specifically, we obtain the
$(3,3,7)$ network configuration: 3 service types, 3 partitions, and 7 slices. The RAN-slicing and PRB-partitioning scenario for this configuration is shown in Figure \ref{fig:network_simulation_slice_partition}, where Partition 1 has two slices: Pre1 and Norm1; Partition 2 has three slices: Pre2, Norm2 and FWA1; and Partition 3 has two slices: Pre3 and FWA2. The unallocated PRBs form the residual partition. The proposed 3-layered framework, \textit{FORSLICE}, for such a hierarchical network structure is also depicted in Figure \ref{fig:network_simulation_FORSLICE}.

In this case study, eMBB Premium slices --- which have higher priority --- are present across all partitions, and every service type has access to more than one partition. We can observe that there is a differentiated quality of service requirements (varied 5QIs in the figure). Fairness in resource allocation implies that each service type receives its required share of PRBs and achieves the desired throughput levels, with a priority to the eMBB Premium service type.

\begin{wrapfigure}{r}{0.5\linewidth}
  \centering
    \includegraphics[scale=0.3]{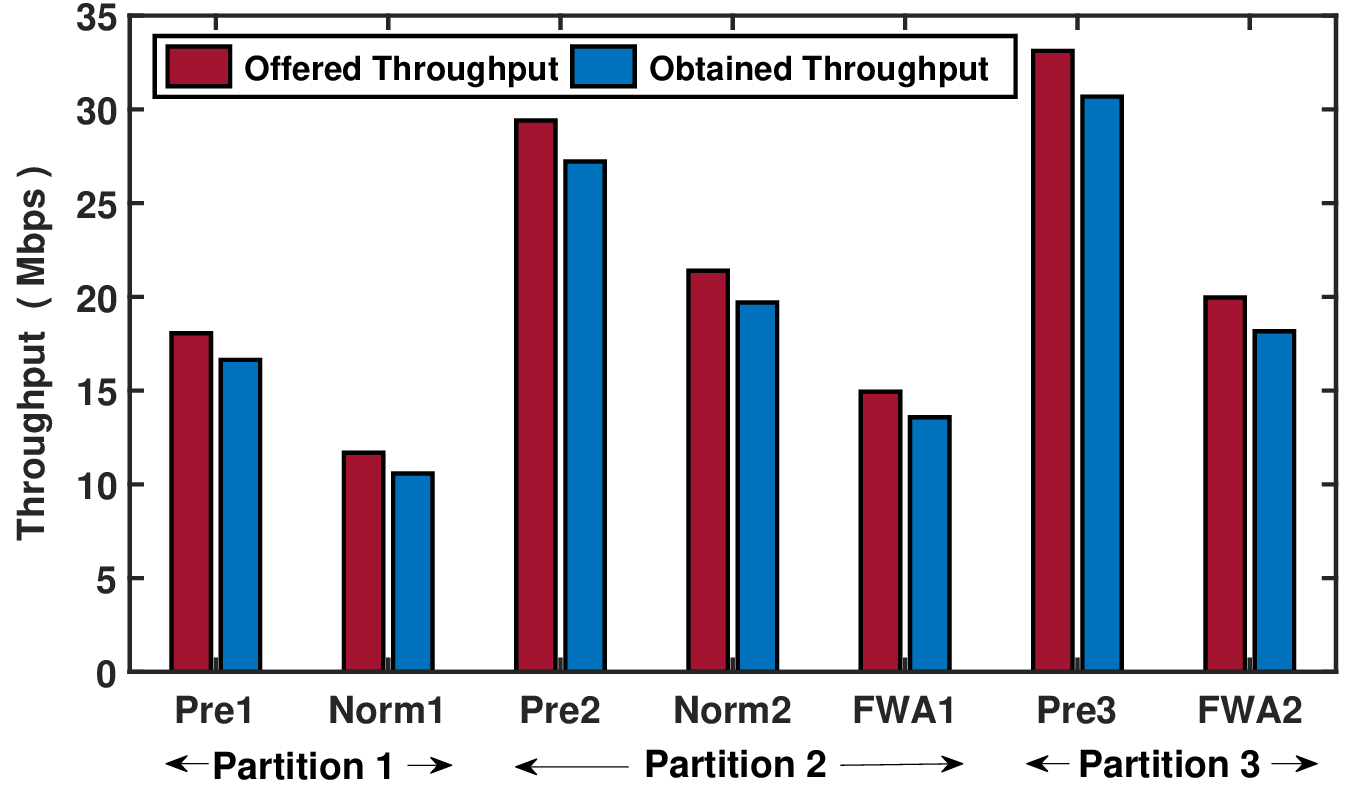}
  \caption{\small Throughput: offered (\textit{FORSLICE}) {\it vs} obtained (NS3-5Glena)}
  \label{fig:throughput}
\end{wrapfigure}

Now, we consider a network simulation for the $(3,3,7)$ configuration,
to demonstrate the network performance of {\it FORSLICE}, measured in terms of the \textit{average throughput} per slice over all the timesteps. Specifically, we show that the throughput achieved by the PRB-allocations and de-allocations in \textit{FORSLICE} is almost similar to the throughput observed in the actual network simulation, indicating that \textit{FORSLICE} maintains the desired network performance.

We simulate the $(3,3,7)$ configuration for $30$ timesteps in the network simulator NS3-5Glena \cite{5G_NS_Glena}.
At each timestep, we associate $1$ to $36$ UEs (based on the user-count in {\it FORSLICE}) with the slices. All the UEs are connected to a single 5G base station (gNB) in a clear Line-of-Sight (LoS) path through an isotropic antenna model \cite{5G_NS_Glena}. Each service type follows a full buffer UDP traffic model \cite{UDP_traffic}. 
Furthermore, we consider the rest of the simulation parameters as the default setup. Specifically, the telecommunication network parameters: central frequency, bandwidth, transmission power, modulation scheme, numerology, and component carrier values are set up at 7GHz, 100MHz, 47dBm, 256QAM, 0, 1, respectively.

We use the PRB-allocation outcome from the {\it FORSLICE} model to calculate the {\it offered throughput}. 
The actual network simulation in NS3-5Glena provides the {\it obtained throughput} for each slice.  Figure \ref{fig:throughput} shows that both the offered and obtained throughput depict nearly similar behaviors for all slices.  
Therefore, we conclude that the allocated PRB-share in each slice attains the throughput as per the expectation (see Eq. \eqref{eq: throughput} and Inference $4$ for the relationship between PRB-usage and throughput requirements). Moreover, the offered and the obtained throughput for Premium service types is always greater than that of the Normal ones, thereby prioritizing the eMBB Premium service type (as observed in Inference $3$).

\section{Related Work}
\label{sec:rwork}
To the best of our knowledge, formal methods have not been applied in the RAN slicing domain in any prior research. 
Thereby, we separately explore the existing methods in two directions: PRB-allocation in RAN and formal methods in telecommunication.

\textbf{PRB-Allocation in RAN}: 
Resource allocation in network slices is well-studied in various contexts \cite{motalleb2022resource,yin2020multiplexing,setayesh2020jointScheduling,convergence}. The work in \cite{motalleb2022resource} uses network slicing to study service-aware baseband resource allocation and virtual network function activation in open-RAN systems. This work deals with the three service types eMBB, URLLC and mMTC, and applies Lagrangian function and Karush-Kuhn-Tucker conditions to obtain optimal PRB-allocation in open-RAN systems. 

The work in \cite{yin2020multiplexing} considers the media access control layer scheduling perspective; analyzes the multiplexing of eMBB and URLLC traffic in 5G downlink transmission. The resource allocation problem, considered for eMBB and URLLC network slices, is formulated as an ILP, such that the original eMBB users' aggregate utility is maximized while guaranteeing the QoS of URLLC users.
The method proposed in \cite{setayesh2020jointScheduling} also accounts for eMBB and URLLC service types, but focuses on joint scheduling of transmit power and PRBs for remote radio heads in the 5G cloud RAN systems. 
The resource-allocation problem is formulated as a mixed-integer non-linear program and a penalized successive convex approximation determines a suboptimal solution in polynomial time.

The only work that considers resource allocation to RAN slices while adopting the similar PRB-partitioning approach as ours, is \cite{convergence}. An AI-planning agent is employed to solve the PRB-allocation problem in \cite{convergence}, accounting for fairness and prioritizing eMBB Premium service. However, their method is limited to a particular case study, whereas our proposed framework is generic and computes PRB-allocation for any configuration of the 3-layered network. Furthermore, the experimental comparisons show that the PRB-allocation in \textit{FORSLICE} exhibits enhanced optimality and also caters the best-effort services, in contrast to \cite{convergence}.

\textbf{Formal Methods in Telecommunication Networks}: In the recent past, formal methods have been used in moderation in the networking area \cite{Formal_ML_networks,Formal_Security,FV_EAP-AKA,idi2022CAC,UML_Uppaal}. 
The authors in \cite{idi2022CAC} address formal modeling and verification of call admission control strategy (CAC), specifically used for voice communication.
Probabilistic model checking and continuous-time Markov Chains model is used to describe the CAC schema. 
The work in \cite{UML_Uppaal} formally verifies an existing 5G service orchestration solution (i.e., coordination of network resources) and checks if the solution meets the service-level agreements over various dynamic behaviors, like, dynamic network load and link utilization. Formal verification is used on a security protocol named, EAP-AKA (Extensible Authentication Protocol - Authenticated Key Agreement) in \cite{FV_EAP-AKA} to guarantee secure transmission of user data. It uses a tool named Pro-Verif, to evaluate the extent to which the attacker can intrude on the communication path. 
Although formal methods explore on different aspects of network problems, it is limited to higher layer resource management. Moreover, the joint exploration on the resource allocation in RAN and formal methodology is still an open area of research.

\section{Conclusion}
\label{sec:conclusion}
In 5G network slicing, \textit{differentiated} quality of service (QoS) is provided to various service types having multiple priorities. 
Hence, to achieve RAN slicing, it is crucial to develop a \textit{dependable} design strategy for resource
allocation to RAN slices, i.e., the system properties like --- ensuring appropriate provision of the desired resources to a higher priority service type, while guaranteeing fairness to all other service types and preserving resource-optimality --- must always be satisfied to uphold network performance standards. 
The basic unit of radio resource is PRB, and the objective of this work is to efficiently allocate PRBs to RAN slices corresponding to multiple service types.  
To ensure the correctness of the solution to the PRB-allocation problem, we formally model the problem as a 3-layered framework, {\it FORSLICE} and we employ formal verification to verify the following system properties: maintenance of a particular range of throughput for Premium users (prioritizing Premium customers), guaranteeing a minimum share of PRBs and throughput to each service (guaranteeing fairness) and minimization of the allocated but unused PRBs (PRB-optimality).
To the best of our knowledge, this is the first work leveraging formal verification in solving the PRB-allocation problem in multi-priority multi-services scenarios. We synthesize an automated formal framework and substantiate the efficiency and applicability of the proposed framework with a handful of experiments. Formal modeling and verification involving a broader range of KPIs offers a compelling avenue for future research.

\begin{acks}
  This work was funded by Ericsson Research India.
\end{acks}

\bibliographystyle{ACM-Reference-Format}
\bibliography{ref}
\end{document}